\documentclass{kapproc} % Computer Modern font calls

\kluwerbib
\let\lcitebracket(
\let\rcitebracket)

\newcommand\paperno{
   \vspace{-30\baselineskip} \hspace{-0.045\textwidth}
   \begin{minipage}[t]{100mm}
   \noindent \small \it Chapter to the book ``Cosmic Gamma-Ray Sources,"\\
                 to be published by Kluwer ASSL Series, \\
                 edited by K.~S.~Cheng and G.~E.~Romero
   \end{minipage}\vspace{27\baselineskip}}
\newcommand{\gray}{$\gamma$-ray}
\newcommand{\grays}{$\gamma$-rays}
\def\gtrsim{\mathrel{\hbox{\rlap{\hbox{\lower3pt\hbox{$\sim$}}}\raise1pt\hbox{$>$}}}}
\def\lsssim{\mathrel{\hbox{\rlap{\hbox{\lower3pt\hbox{$\sim$}}}\raise1pt\hbox{$<$}}}}

\newcommand{\pubjournal}[5]{#1, {\bf #2}, #3, #4.}
\newcommand{\pubproc}[4]{#1, p.#2, #3.}
\newcommand{\araa}{Annual Rev.\ Astron.\ Astrophys.}
\newcommand{\aap}{Astron.\ Astrophys.}
\newcommand{\aaps}{Astron.\ Astrophys.\ Suppl.}
\newcommand{\adv}{Adv.\ Space Res.}
\newcommand{\app}{Astropart.\ Phys.}
\newcommand{\apj}{Astrophys.\ J.}
\newcommand{\apjl}{Astrophys.\ J.}

\newcommand{\mnras}{Mon.\ Not.\ Royal Astron.\ Soc.}
\newcommand{\nat}{Nature}

\newcommand{\prd}{Phys.\ Rev.\ D}
\newcommand{\prl}{Phys.\ Rev.\ Lett.}
\newcommand{\ssr}{Spa.\ Sci.\ Rev.}
\newcommand{\icrc}{Int.\ Cosmic Ray Conf.}
\newcommand{\citep}{\cite}
\newcommand{\citet}{\cite}
\begin{document}

%\articletitle{Galactic and Extragalactic Diffuse Gamma Rays}
\articletitle{Diffuse Gamma Rays}
\articlesubtitle{Galactic and Extragalactic Diffuse Emission}

\author{Igor V.~Moskalenko\footnote{JCA/University of 
Maryland, Baltimore County, Baltimore, MD 21250, USA}}
\affil{NASA/Goddard Space Flight Center, Code 661\\
Greenbelt, MD 20771, USA}
%and\\
%JCA/University of Maryland, Baltimore County,\\
%Baltimore, MD 21250, USA}
\email{Igor.Moskalenko@gsfc.nasa.gov}

\author{Andrew W.~Strong}
\affil{Max-Planck-Institut f\"ur extraterrestrishe Physik\\
Postfach 1312\\
85741 Garching, Germany}
\email{aws@mpe.mpg.de}

\author{Olaf Reimer}
\affil{Ruhr-Universit\"at Bochum\\
Theoretische Physik, Lehrstuhl IV,
Weltraum- und Astrophysik\\
44780 Bochum, Germany}
\email{olr@tp4.rub.de}

\paperno

\begin{abstract}
%######################################################################

``Diffuse'' gamma rays consist of several components:  truly diffuse
emission from the interstellar medium, the extragalactic background,
whose origin is not firmly established yet, and the contribution from
unresolved and faint Galactic point sources.  One approach to unravel
these components is to study the diffuse emission from the
interstellar medium, which traces the interactions of high energy
particles with interstellar gas and radiation fields. Because of its
origin such emission is potentially able to reveal much about the
sources and propagation of cosmic rays. The extragalactic background,
if reliably determined, can be used in cosmological and blazar
studies.  Studying the derived ``average'' spectrum of faint Galactic
sources may be able to give a clue to the nature of the emitting
objects.

\end{abstract}

%\begin{keywords}
%Diffuse Galactic continuum gamma rays, diffuse extragalactic gamma rays,
%cosmic rays, Galactic structure
%\end{keywords}

\section*{Introduction}
%######################################################################

As is discussed in detail in the first Chapter of this book, the
subject of \gray\ astronomy was born in 1972 when the first
statistically significant results were obtained by the SAS-2
satellite.  This was followed by the COS-B observatory in 1975--1982
and several low-energy missions. The Compton Gamma-Ray Observatory
(CGRO) launched in 1990 had 4 instruments onboard covering the energy
range from 20 keV to 30 GeV and was very  successful. Besides many
observations of point sources, COS-B and then COMPTEL and EGRET (two
of four  CGRO instruments) unveiled the spectrum of the diffuse
Galactic continuum emission and thus have shown the potential of
\gray\ observations to contribute to cosmic ray physics.

The diffuse \gray\ emission supposedly consists of several components:
truly diffuse Galactic emission from the interstellar medium, the
extragalactic background, whose origin is not firmly established yet,
and the contribution from unresolved and faint Galactic point sources.
The Galactic diffuse emission dominates other components and  has a
wide distribution with most emission coming from the Galactic plane.

Diffuse continuum \grays\ from the interstellar medium are
potentially able to reveal much about the sources and propagation of
cosmic rays, but in practice the exploitation of this well-known
connection is not straightforward.  The Galactic diffuse continuum
\grays\ are produced in energetic interactions of nucleons with gas
via neutral pion  production, and by electrons via inverse Compton
scattering and bremsstrahlung.  These processes are dominant in
different parts of the spectrum, and therefore if deciphered the
\gray\ spectrum can provide information about the large-scale spectra
of nucleonic and leptonic components of cosmic rays. In turn, having
an improved understanding of the Galactic diffuse \gray\ emission and
the role of cosmic rays is essential for unveiling the spectra of
other components of the diffuse emission and is thus of critical
importance   for the study of many topics in \gray\ astronomy, both
Galactic and extragalactic.

The launch in 2007 of the Gamma-ray Large Area Space Telescope (GLAST)
will tremendously increase the quality and accuracy of the \gray\
data. Study of the diffuse emission is one of its priority goals.  In
the present Chapter, we concentrate on the high energy ($E>100$ MeV)
part of the diffuse emission.

\section{Gamma Rays and Cosmic Rays Connection} \label{connection}
%######################################################################

The Galactic diffuse \gray\ continuum emission, which arises from
cosmic-ray proton and electron interactions with gas and interstellar
radiation fields, is the dominant feature of the \gray\ sky.  This
emission in the range 50 keV -- 50 GeV has been systematically studied
in the experiments OSSE, COMPTEL, EGRET on the CGRO as well as in
earlier experiments, such as SAS 2 and COS B.  A review of CGRO
observations was presented by Hunter et al.\ (1997).

\begin{figure}[t]%%%%%%%%%%%%%%%%%%%%%%%%%%%%%%%%%%%%%%%%%%%%%%%%%%%%%
\vskip 2.65in 
\includegraphics{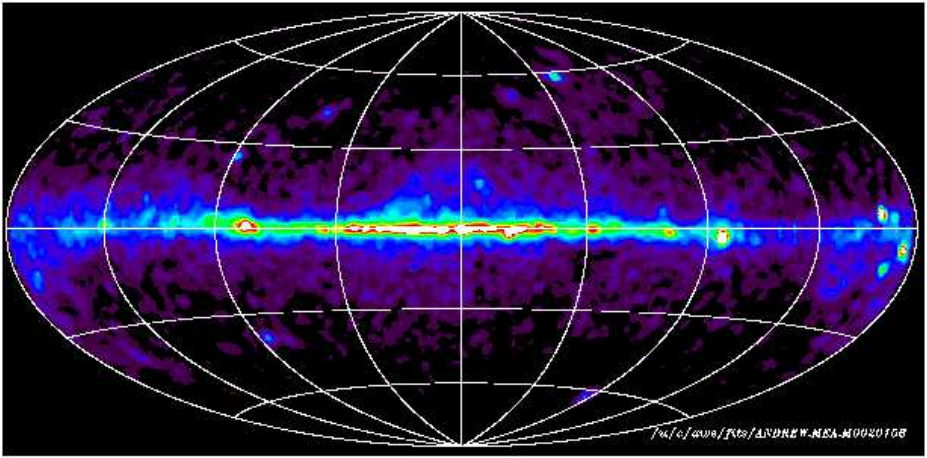}
\caption{EGRET all-sky map in continuum \gray\ emission for energies 
$>$100 MeV (A.~W.~Strong, unpublished). \label{fig:skymap}}
\end{figure}%%%%%%%%%%%%%%%%%%%%%%%%%%%%%%%%%%%%%%%%%%%%%%%%%%%%%%%%%%

The great sensivity and spatial and energy resolution of the EGRET
instrument allowed for detailed spatial and spectral analysis  of the
diffuse emission (Fig.\ \ref{fig:skymap}). Because the  Galaxy is
transparent to high energy \grays, the diffuse \gray\ emission is the
line-of-sight integral over the emissivity of the interstellar
medium. The latter is  essentially the product of the cosmic ray
density and the density of the gas or radiation field. The hydrogen
distribution (H$_2$, H {\sc i}, H {\sc ii}) is derived from radio
surveys and an assumed Galactic rotation curve, where the distribution
of molecular hydrogen is derived indirectly from CO radio-emission and
the assumption that the conversion factor H$_2$/CO is the same for the
whole Galaxy.  The Galactic radiation field consists of contributions
of stars, dust, and cosmic microwave background (CMB). Its spectrum
varies  over the Galaxy and (apart from the CMB) cannot be measured
directly.

The first detailed analysis of the diffuse emission from the plane
$|b|\leq10^\circ$ was made by Hunter et al.\ (1997).  The basic
assumptions of this calculation were that  (i) the cosmic rays are
Galactic in origin, (ii) a correlation exists between the cosmic ray
density and interstellar matter in the Galaxy, and (iii) that the
spectra of nucleons and electrons in the Galaxy are the same as
observed in the solar vicinity.  This analysis confirmed results of
earlier experiments \citep{knifen73,fichtel75,mayer82} that the great
majority of the emission is clearly correlated with  the
\emph{expected} Galactic diffuse emission.  It was also shown
\citep{strong88} that, on average, there is a generally decreasing
\gray\ emissivity per H atom, and hence a decreasing cosmic ray
density, with Galactic radius.

\begin{figure}[t]%%%%%%%%%%%%%%%%%%%%%%%%%%%%%%%%%%%%%%%%%%%%%%%%%%%%%
\vskip 2.05in 
%\special{psfile=diffuse_f2.ps voffset=-200 hoffset=-10 vscale=70 hscale=90} 
\includegraphics{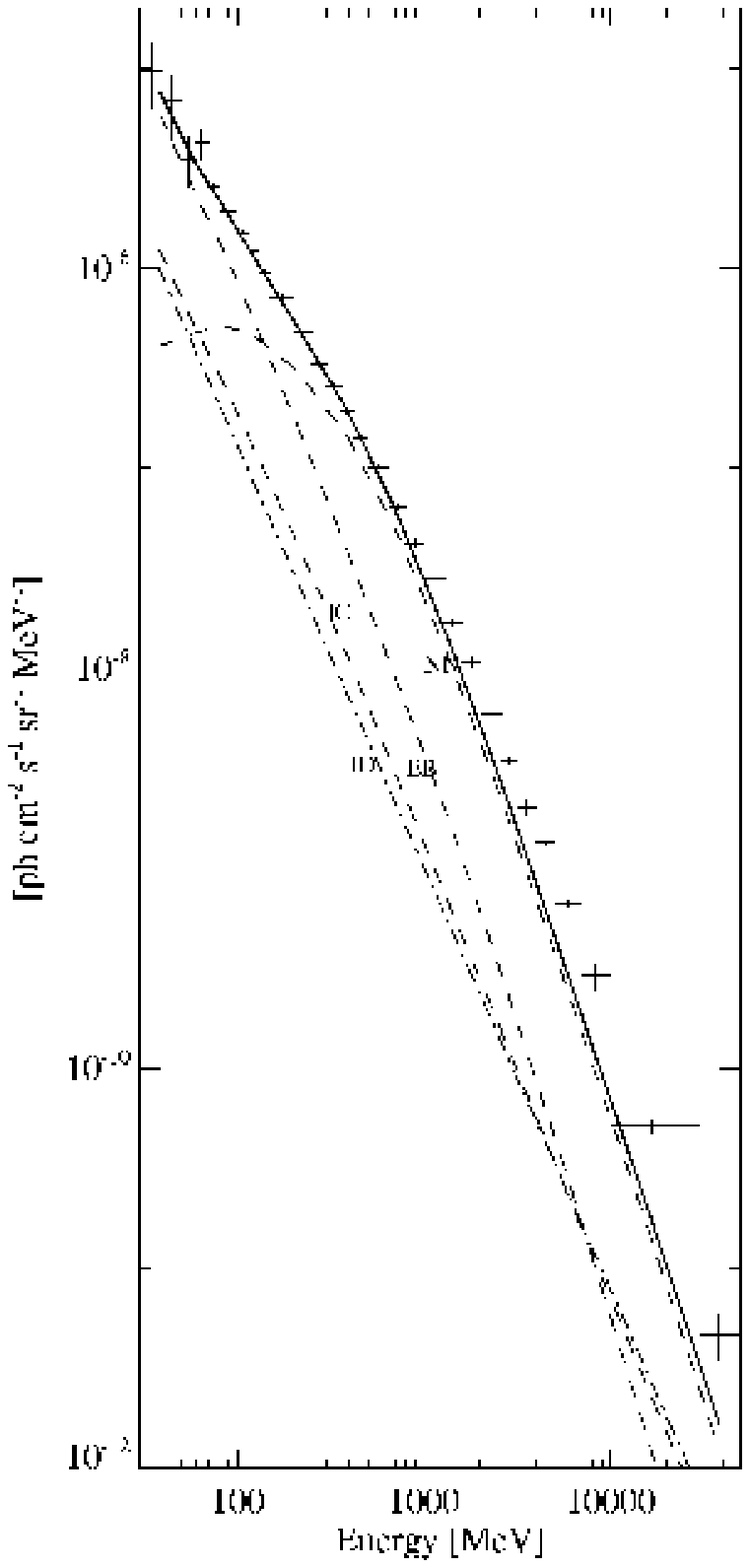} 
\narrowcaption{Average diffuse gamma-ray spectrum
of the inner Galaxy region, $300^\circ<l<60^\circ, |b|\le10^\circ$.
The contribution from point sources detected with more than $5\sigma$
significance have been removed. The individual components of this
calculation are nucleon-nucleon (NN), electron bremsstrahlung (EB),
inverse Compton (IC), and isotropic diffuse emission (ID). Adapted
from Hunter et al.\ (1997). \label{fig:hunter97}}
\end{figure}

The observations have confirmed main features of the Galactic model
derived from cosmic rays, however, they brought also new puzzles.  The
\grays\ revealed that the cosmic ray source distribution required to
match the \gray\ data apparently should be distinctly flatter
\citep{strong96} than the (poorly) known distribution of supernova
remnants (SNRs), the conventional sources of cosmic rays. The spectrum
of \grays\ calculated under the assumption that the proton and
electron spectra in the Galaxy resemble those measured locally reveals
an excess at $>$1 GeV in the EGRET spectrum (Fig.\
\ref{fig:hunter97}).

%% To make narrow caption:
%%%%%%%%%%%%%%%%%%%%%%%%%%%%%%%%%%%%%%%%%%%%%%%%%%%%%%%%%

The puzzle of the ``GeV excess'' has lead to an attempt to
re-evaluate the reaction of $\pi^0$-production in
$pp$-interactions. However, a calculation \citep{mori97} made using
modern Monte Carlo event generators to simulate high-energy
$pp$-collisions has shown that the \gray\ flux agrees rather well with
previous calculations.

Leaving the possibility of a instrumental artefact aside, another
leading reason for the discrepancy discussed is that the local cosmic
ray particle spectra (nucleons and/or electrons) may be not
representative of the Galactic average.  The local source(s) and
propagation effects (e.g., electron energy losses) can change the
spectrum of accelerated particles.

A flatter Galactic nucleon spectrum has been suggested as a possible
solution to the ``GeV excess'' problem
\citep{mori97,gralewicz97}. Explaining the excess requires the
power-law index of proton spectrum of about --2.4--2.5.  A flatter
electron spectrum has been proposed by Porter and Protheroe (1997) and
Pohl and Esposito (1998).  The average interstellar electron spectrum
can be harder than that locally observed due to the spatially
inhomogeneous source distribution and energy losses.  The \gray\
excess in this case may be explained in terms of inverse Compton
emission.

However, the average energy spectrum of the diffuse \gray\ emission
alone does not tell much about the underlying processes.  Instead, the
spectrum of diffuse continuum \gray\ emission  from different
directions and its distribution on the sky carry unique information
about the particle fluxes, mostly protons and electrons, in different
locations.   The $\pi^0$-decay and bremsstrahlung photons are gas
related and thus should mimic the distribution of interstellar matter.
In contrast, inverse Compton emission is broad since the density of
background photons is high at even large distances from the Galactic
plane.  In practice, however, this simple picture is complicated
because the H\,{\sc ii} gas distribution is broad with typical scale
height $\sim$1 kpc, while the distribution of high energy electrons is
narrow and concentrated near the Galactic plane due to the large
energy losses.  To decode the wealth of information provided to us
by diffuse \grays\ from different directions one needs a proper
propagation model to calculate the particle spectra and the corresponding
\gray\ flux on a large Galactic scale.

A self-consistent model of particle propagation and generation of
diffuse \grays\ should include cosmic-ray transport as the first
step. Knowing the number density of primary nuclei from satellite and
balloon observations, the production cross sections from 
laboratory experiments, and the gas distribution from astronomical
observations, one can calculate the production rate of secondary
nuclei. The observed abundance of radioactive isotopes determines then
the value of the diffusion coefficient, halo size and other global
parameters.  The detailed procedure was described, e.g., by Ptuskin
and Soutoul (1998) and Strong and  Moskalenko (1998). Having fixed the
propagation model  and assuming some particle spectra in the
cosmic-ray sources, this allows one to calculate the spectrum of the
diffuse  \gray\ emission \citep{strong00}.

\section{Cosmic Rays} \label{cosmicrays}
%######################################################################

Cosmic rays are energetic particles, which come to us from outer
space, and are measured either with satellites, balloons, or Earth
based experiments. Direct measurements give the spectra in the local
region of the Galaxy. For energies below 10 GeV the heliospheric
modulation is large and this hinders the study of  the truly
interstellar spectrum.

The spectrum of cosmic rays can be approximately described by a single
power law with index $-3$ from $\sim$10 GeV to the highest energies
ever observed $\sim$$10^{20}$ eV. The only feature
observed below $10^{18}$ eV is a small change in the slope from 
--2.7 to --3.1 at $\sim3\times10^{15}$ eV, known as the ``knee.'' 
Because of this featureless
spectrum, it is believed that cosmic-ray production and propagation is
governed by the same mechanism over decades of energy; a single
mechanism at least works below the knee, at $\sim$$10^{15}$ eV,  and
the same or another one works above the knee. Meanwhile the origin of
the cosmic-ray spectrum is not still understood.

Galactic cosmic rays are an important part of the interstellar medium.
The energy density of relativistic particles is about 1 eV cm$^{-3}$
and is comparable to the energy density of the interstellar radiation
field, magnetic field, and turbulent motions of the interstellar
gas. This makes cosmic rays one of the essential factors determining
the dynamics and processes in the interstellar medium.  The EGRET
observations of the Small Magellanic Cloud  \citep{sreekumar93} have
shown that the cosmic rays are  a Galactic and \emph{not} a
``metagalactic'' phenomenon. Observations of the Large Magellanic
Cloud \citep{sreekumar92}, on the other  hand, have shown that \gray\
emission is consistent with quasi-static equilibrium of cosmic rays
and the interstellar medium.
 
\begin{figure}[t]%%%%%%%%%%%%%%%%%%%%%%%%%%%%%%%%%%%%%%%%%%%%%%%%%%%%%
\vskip 3.0in 
\includegraphics{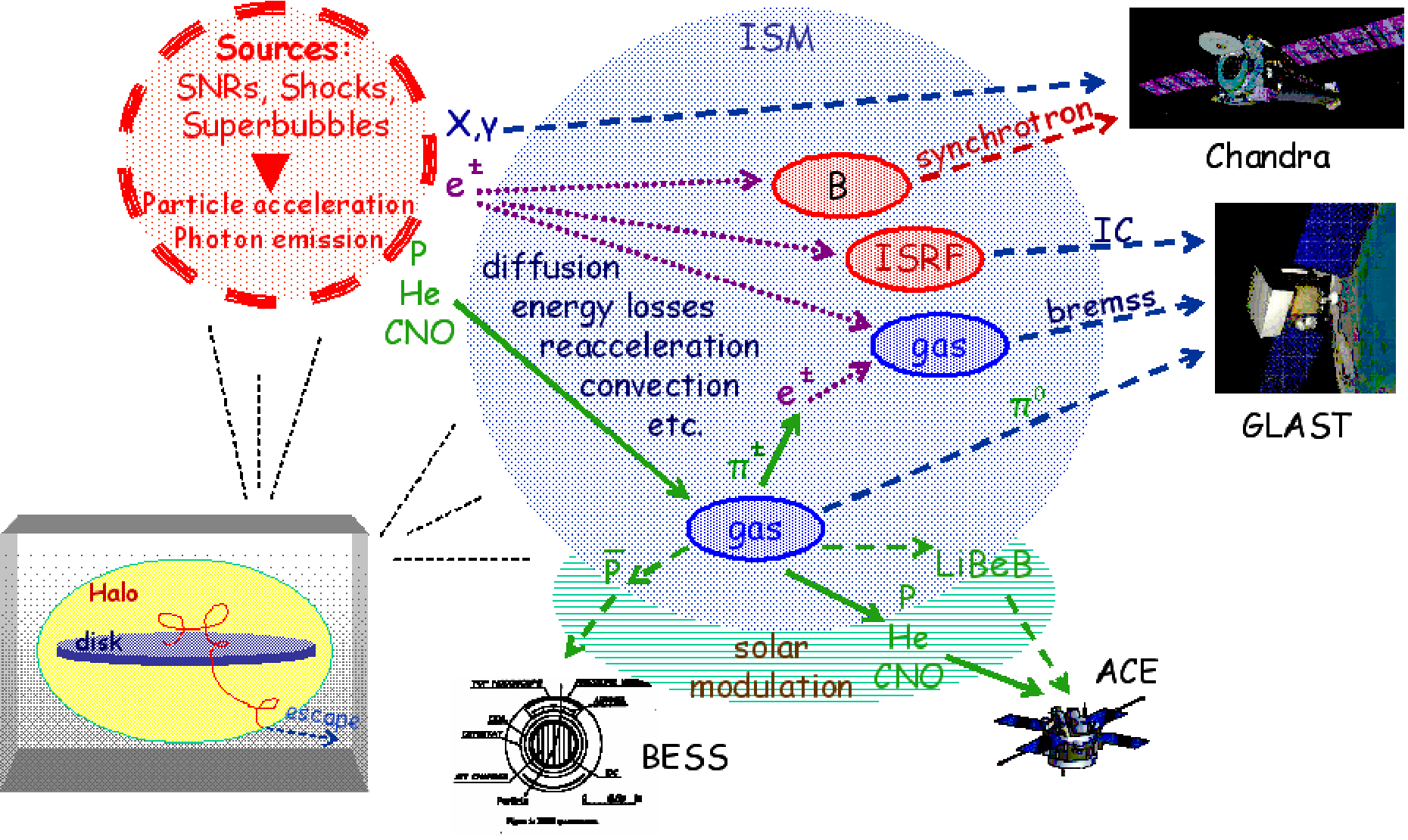}
\caption{A schematic view of cosmic ray propagation in the
interstellar medium (ISM), production of secondary nuclei, particles
and \grays. \label{fig:cr_propagation}}
\end{figure}%%%%%%%%%%%%%%%%%%%%%%%%%%%%%%%%%%%%%%%%%%%%%%%%%%%%%%%%%%

The sources of cosmic rays are believed to be supernovae and SNRs,
pulsars, compact objects in close binary systems, and stellar winds.
Observations of X-ray and \gray\ emission from these objects reveal
the presence of energetic particles thus testifying to efficient
acceleration processes near these objects.  The total power of
Galactic cosmic ray sources needed to sustain the observed cosmic ray
density is estimated at $5\times10^{40}$ erg s$^{-1}$ which implies
the release of energy in the form of cosmic rays of approximately
$5\times10^{49}$ erg per supernovae if the supernova rate in the
Galaxy is 1 every 30 years. This value comes to about 5\% of the
kinetic energy of the ejecta which is in agreement with the prediction
of the theory of diffusive shock acceleration \citep{jones91}. This
scenario implies that cosmic rays accelerated by the shock waves
propagate further in the Galaxy where they are contained for some 10
Mys before escaping into intergalactic space.

Particles accelerated near the sources propagate in the interstellar
medium  (Fig.\ \ref{fig:cr_propagation}) where they lose or gain
energy, their initial spectra and composition change, they produce
secondary particles and \grays. The destruction of primary nuclei via
spallation gives rise to secondary nuclei and isotopes which are rare
in nature, antiprotons, and pions  which decay producing $\gamma$-rays
and secondary positrons and electrons. Because secondary antiprotons,
positrons, and diffuse \grays\ (via neutral pion decay) are all
products of the same $pp$-interactions,  accurate measurements of
the antiproton and positron fluxes, especially at high energies, could
provide a diagnostic of the interstellar nucleon spectrum
complementary to that provided by \grays\
\citep{moskalenko98,strong00}.

The variety of isotopes in cosmic rays allow one to study different
aspects of their acceleration and propagation in the interstellar
medium as well as the source composition. Stable secondary nuclei tell
us about the diffusion coefficient and Galactic winds (convection)
and/or re-acceler\-a\-tion in the interstellar medium (2nd order Fermi
acceleration mechanism).  Long-lived radioactive secondaries allow one
to constrain global Galactic properties such as, e.g., Galactic halo
size.  Abundances of K-capture isotopes,  which decay via electron
K-capture after attaching an electron from the ISM, can be used to
probe the gas density and acceleration time scale.  All these together
allow us in principle to build a model of particle acceleration and
propagation in the Galaxy.

The most often used propagation model, the flat halo diffusion model, has a simple
geometry which reflects however the most essential features of the
real system \citep{ginzburg}. It is assumed that the Galaxy has the
shape of a cylinder with a radius $R$ ($\sim$20 kpc) and total height
$2H$ ($H>1$ kpc). The cosmic-ray sources are distributed within an
inner disk having characteristic thickness $\sim$300 pc. The Sun is at
a distance $\sim$8 kpc from the center of the Galaxy. The diffusion of
cosmic rays averaged over the scale of few hundred parsec is
isotropic. The particles escape freely through the halo boundaries
into intergalactic space where the density of cosmic rays is
negligible.

The modelling of cosmic-ray diffusion in the Galaxy includes the
solution of the transport equation with a given source distribution
and boundary conditions for all cosmic-ray species. The transport
equation describes diffusion, convection by the hypothetical Galactic
wind, energy losses, and possible distributed acceleration (energy
gain).  The study of transport of cosmic-ray nuclear component
requires the consideration of nuclear spallation and ionization energy
losses. Calculation of isotopic abundances is impossible without
inclusion of hundreds of stable and radioactive isotopes produced in
the course of cosmic-ray interactions with interstellar gas.

%% To make narrow caption:
\begin{figure}[t]%%%%%%%%%%%%%%%%%%%%%%%%%%%%%%%%%%%%%%%%%%%%%%%%%%%%%
\vskip 1.85in  
%\special{psfile=diffuse_f4.ps voffset=-120 hoffset=-20 vscale=90 hscale=90}  
\includegraphics{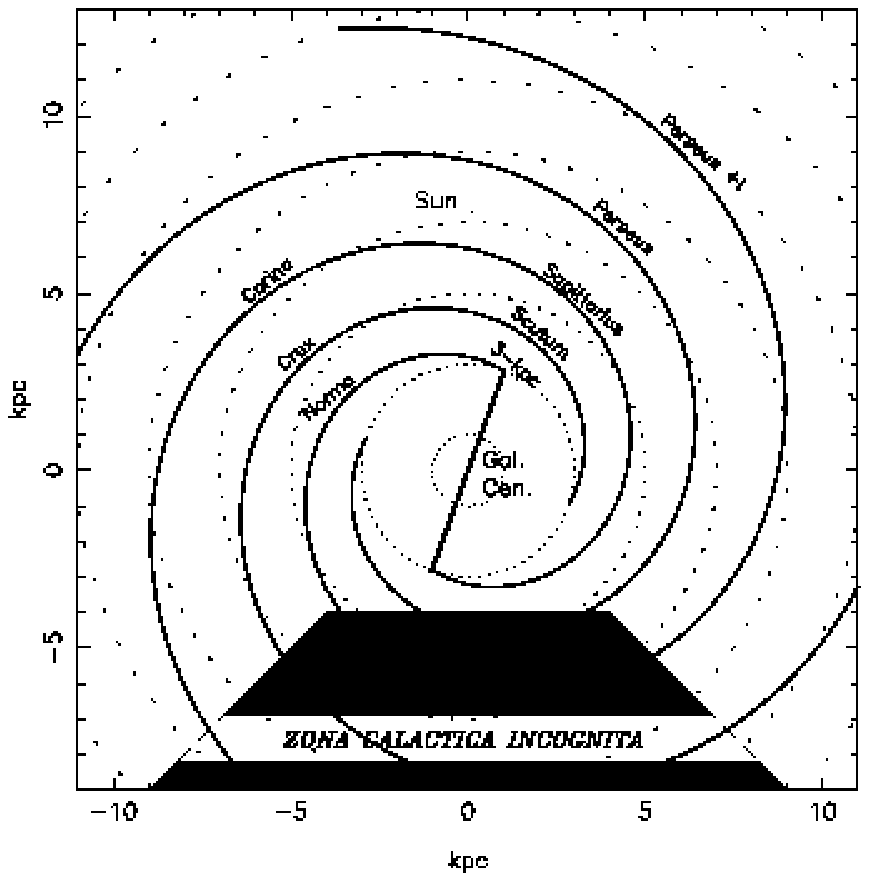}  
\narrowcaption{Model of logarithmic spiral arms.
The sun is shown by the circled dot. Dots show the concentric circles
at the Galactocentric radii 1, 3, 5, 7, 9, 11, 13 kpc.  Adapted from
Vall\'ee (2002). \label{fig:vallee02}}
\end{figure}%%%%%%%%%%%%%%%%%%%%%%%%%%%%%%%%%%%%%%%%%%%%%%%%%%%%%%%%%%

\section{Galactic Structure} \label{structure}
%######################################################################

The Galaxy is a barred spiral with a radius of about 30 kpc  (Fig.\
\ref{fig:vallee02}).  From the point of view of \gray\ diffuse
emission the important components are the gas and the interstellar
radiation, while synchrotron emission provides
restrictions on the electron spectrum. These are also relevant for the
energy losses of cosmic rays.  The gas content is dominated by atomic
(H\,{\sc i}) and molecular hydrogen (H$_2$),  which are present in
approximately equal quantities  ($\sim$$10^{9}\ M_{\odot}$) in the
inner  Galaxy, but with very different radial distributions.  There is
also a small fraction of low-density ionized hydrogen (H\,{\sc ii}).
In addition to hydrogen, the interstellar gas contains heavier
elements, dominated by helium, with a ratio of $\sim$10\% by number
relative to hydrogen. Helium is therefore an important contributor to
the gas-related \gray\ emission.

\subsection{Interstellar Gas}
%######################################################################

\begin{figure}[t]%%%%%%%%%%%%%%%%%%%%%%%%%%%%%%%%%%%%%%%%%%%%%%%%%%%%%
\vskip 1.25in
\includegraphics{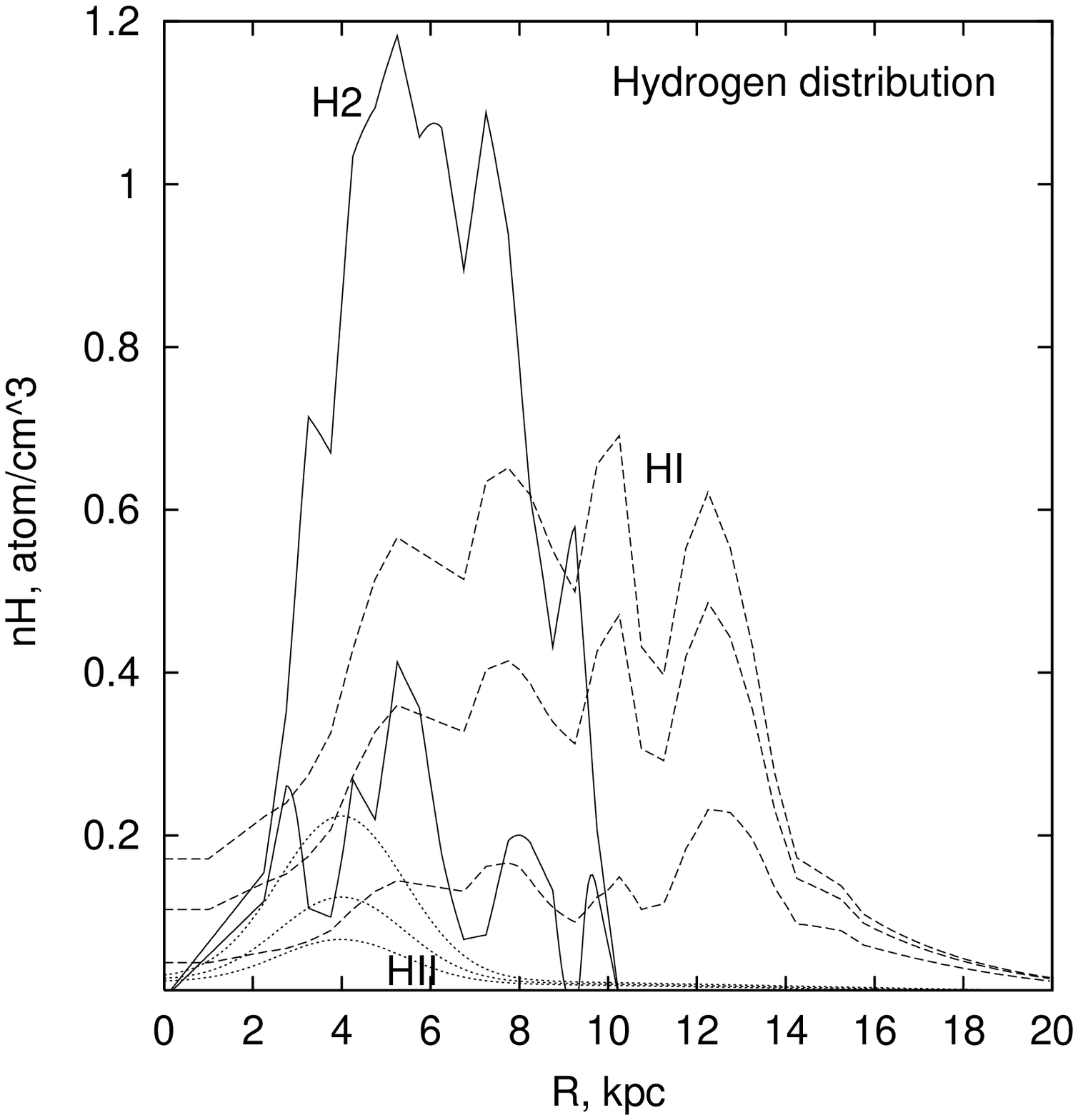}
\narrowcaption{Number density distributions of $2\times$H$_2$
(solid), H\ {\sc i} (dashes), and H\ {\sc ii} (dots) in the Galaxy. Shown are the
plots for $z=0, 0.1, 0.2$ kpc (decreasing density). Number density of
H$_2$ at $z=0.2$ kpc from the plane is very low and is not shown in
the plot. Adapted from Moskalenko et al.\ (2002). \label{fig:hydrogen}}
\end{figure}

The molecular hydrogen H$_2$ is distributed within $R<10$ kpc, with a
peak around 5 kpc and a small scale height, about 70 pc (Fig.\
\ref{fig:hydrogen}). It is concentrated mainly in dense clouds of
typical density $10^{4}$ atom cm$^{-3}$ and masses $10^4-10^6
M_\odot$. The H$_2$ gas cannot be detected directly on large scales,
but the 115 GHz emission of the abundant molecule $^{12}$CO is a good
``tracer,''  since it forms in the dense clouds where the H$_2$
resides.  The derivation of H$_2$ density from the CO data is
problematic; normally a linear relation is assumed and the conversion
factor is derived from independent estimates of the mass of gas,
including the assumption of virial equilibrium, and \gray\ analyses.
The recent result obtained from a complete CO survey and infrared and
H\ {\sc i} maps gives average $X\equiv N_{{\rm H}_2}/W_{\rm CO}=
1.8\times10^{20}$ cm$^{-2}$ K$^{-1}$ km$^{-1}$ s \citep{dame01}.  The
\gray\ method has the advantage of sampling large regions of the
Galaxy and requiring only the assumption that cosmic rays freely
penetrate molecular clouds. An analysis of EGRET sky survey yields
$X=(1.9\pm0.2)\times10^{20}$ cm$^{-2}$ K$^{-1}$ km$^{-1}$ s for
$E_\gamma=0.1-10$ GeV \citep{strong96} without significant energy
dependence, consistent with earlier COS-B analysis
\citep{strong88}. Observations of particular local clouds
\citep{digel96,digel99,digel01,hunter94}  yield somewhat lower values
$X=(0.9\div1.65)\times10^{20}$ cm$^{-2}$ K$^{-1}$ km$^{-1}$ s and
error bars 15--20\%, but still close to the average.  In the outer
Galaxy $X$ may increase.  A simple parametrization of H$_2$
distribution is given in Bronfman et al.\ (1988).

%% To make narrow caption:
%%%%%%%%%%%%%%%%%%%%%%%%%%%%%%%%%%%%%%%%%%%%%%%%%%%%%%%%%

The atomic gas extends out to 30 kpc, with surface density  increasing
with distance from the Galactic center from $1.9 M_\odot$ pc$^{-2}$
within $R=6$ kpc to $\sim$$4 M_\odot$ pc$^{-2}$ at 7--12 kpc, and then
decreasing to $\sim$$1 M_\odot$ pc$^{-2}$ at 17 kpc \citep{nakanishi03}.
The H\,{\sc i} disk is asymmetric with warping in the outer disk, and
it extends to about 1.5 kpc above the Galactic plane in the northern
hemisphere and down to about 1 kpc in the southern hemisphere.  The
gas density is roughly uniform at 1 atom cm$^{-3}$ and a  typical
scale height is about 200 pc.  H\,{\sc i} gas is mapped directly via
its 21 cm radio line,  which gives both distance (from the
Doppler-shifted velocity  and Galactic rotation models) and density
information.  Less studied is a cold component of H\,{\sc i}, which
does not emit at 21 cm. Its presence is detected using absorption
spectra measured against bright extragalactic radio sources. A study
\citep{kolpak02} shows a clear correlation the with H$_2$ distribution.
A simple parametrization of the H\,{\sc i} distribution can be found in
Gordon and Burton (1976) and Dickey and Lockman (1990).

Ionized hydrogen H\,{\sc ii} is present at lower densities, but with
much larger vertical extent.  The ``warm ionized medium'' has
densities $\sim$$10^{-3}$ atom cm$^{-3}$ and a scale height of 1
kpc. This gas makes a small contribution to the \gray\ emission, but
is nevertheless of interest because it produces a much broader
latitude distribution than the neutral gas.  A simple parametrization
of H\,{\sc ii} distribution can be found in Cordes et al.\ (1991).

%% Double captions:
\begin{figure}[t]%%%%%%%%%%%%%%%%%%%%%%%%%%%%%%%%%%%%%%%%%%%%%%%%%%%%%
\vskip 2.55in
\includegraphics{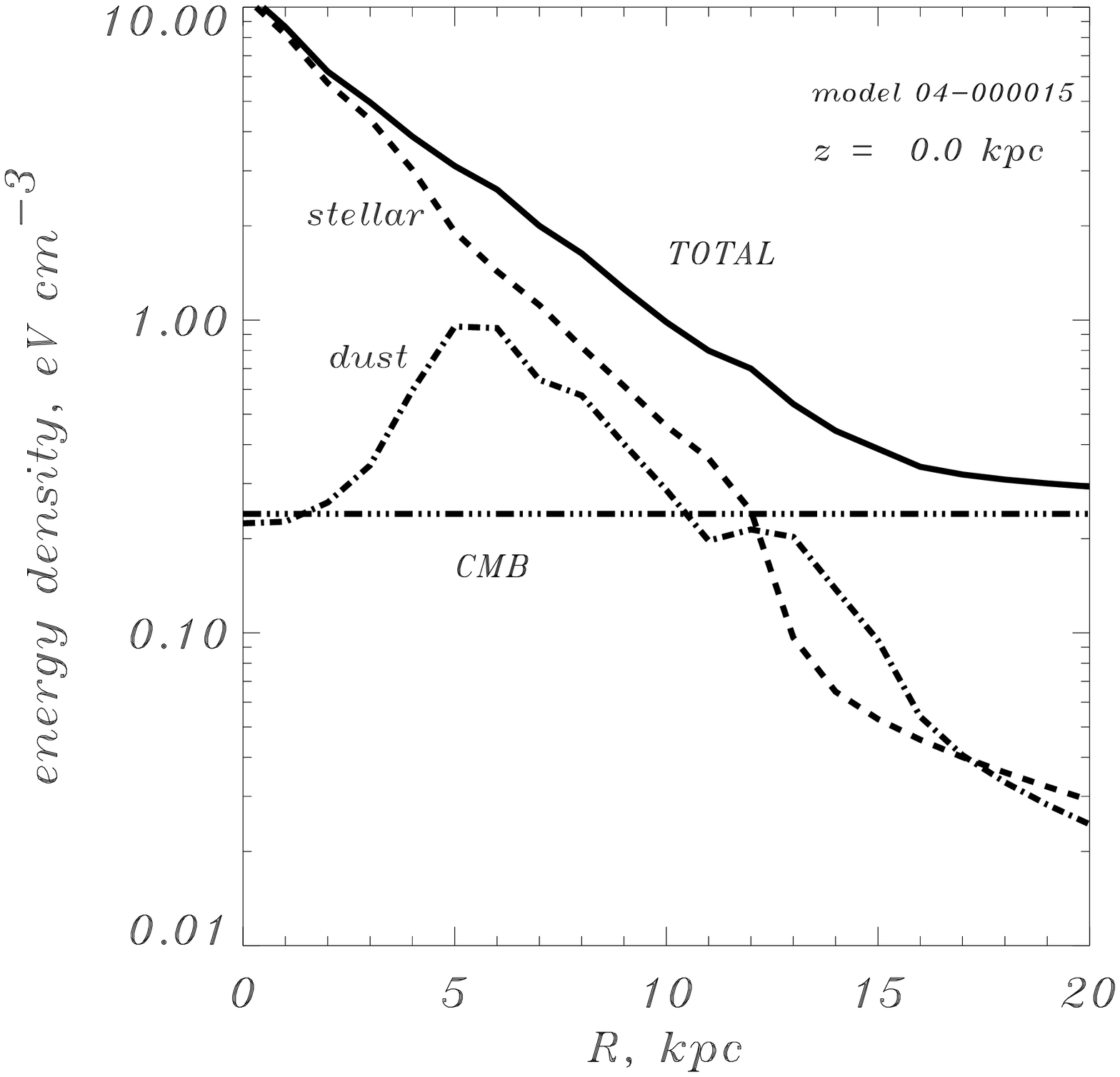}
\includegraphics{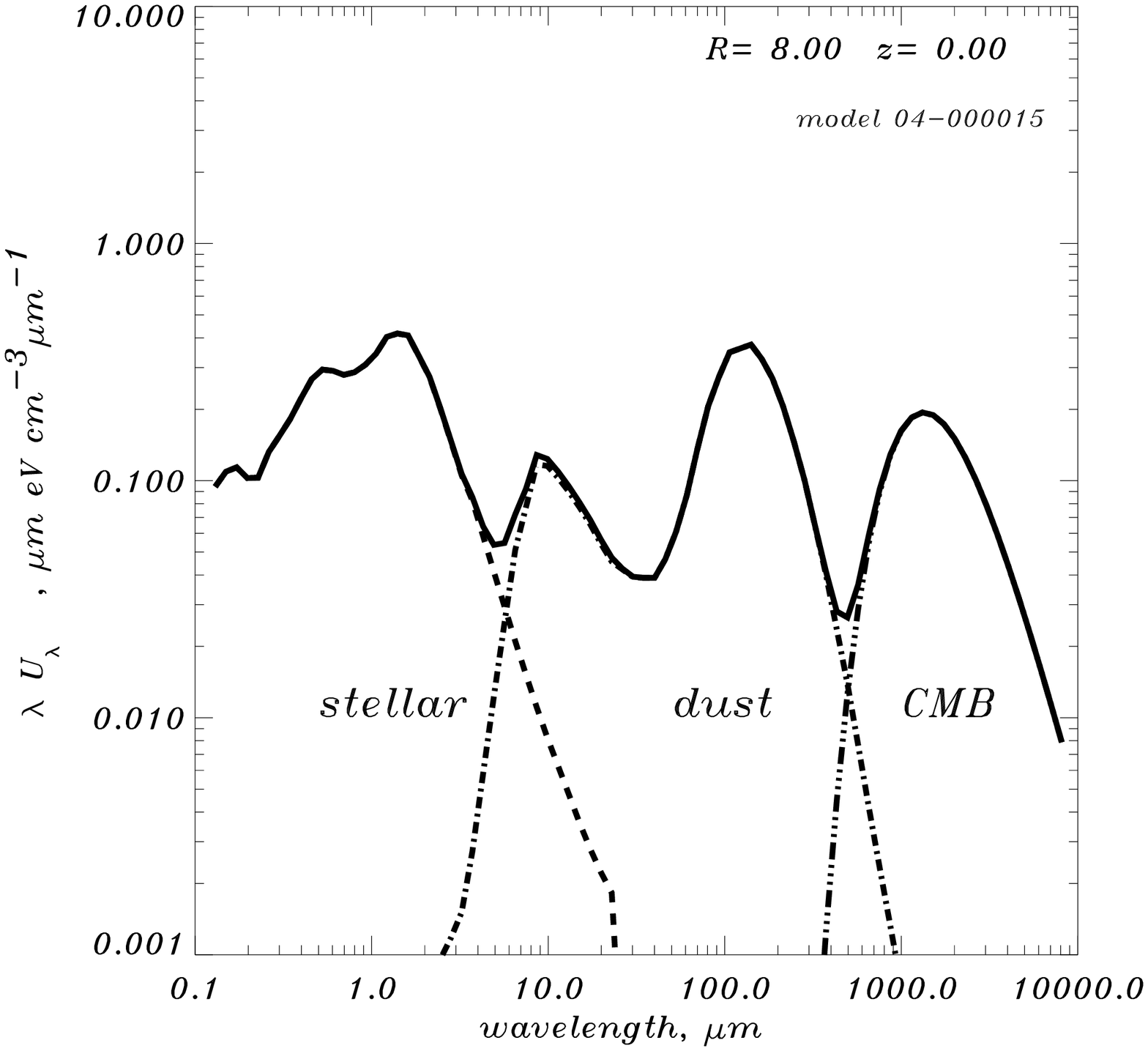}
\caption{\emph{Left:} ISRF energy density as function of $R$ at
$z=0$. \emph{Right:}  The spectrum of the ISRF at $R=8$ kpc,
$z=0$. Adapted from Strong et al. (2000). \label{fig:ISRF}}
\end{figure}%%%%%%%%%%%%%%%%%%%%%%%%%%%%%%%%%%%%%%%%%%%%%%%%%%%%%%%%%%

\subsection{Interstellar Radiation Field}
%######################################################################

The interstellar radiation field (ISRF) is essential for electron
propagation  (energy losses) and \gray\ production by inverse Compton
emission. It is made up of contributions from starlight, emission from
dust, and the CMB.  Estimation of the spectral and spatial
distribution of the ISRF relies on models of the distribution of
stars, absorption, dust emission spectra and emissivities  and is
therefore in itself a complex subject.

New data from infrared surveys by the IRAS and COBE  (Cosmic
Background Explorer) satellites have greatly improved our knowledge of
both the stellar distribution and the dust emission.  Fig.\
\ref{fig:ISRF} (left) shows ISRF energy density as function of
Galactocentric radius, and  Fig.\ \ref{fig:ISRF} (right) shows a
recent estimate of the spectrum at $R=8$ kpc, near the solar
position.  Stellar emission dominates from 0.1 $\mu$m to 10 $\mu$m,
and emission from very small dust grains contributes from  10 $\mu$m
to 30 $\mu$m.  Emission from dust at $T\sim20$ K dominates from 20
$\mu$m to 300 $\mu$m.  The 2.7 K microwave background is the main
radiation field above 1000 $\mu$m.  The ISRF has a vertical extent of
several kpc, where the Galaxy acts as a disk-like source of radius
$\sim$10 kpc. The radial distribution of the stellar component is also
centrally peaked, since the stellar density increases exponentially
inwards with a scale-length of $\sim$2.5 kpc until the bar is
reached. The dust component is related to that of the neutral gas
(H\,{\sc i} + H$_2$) and is therefore distributed more uniformly in
radius than the stellar component.

\subsection{Magnetic Field and Synchrotron Emission}\label{synch_emis}
%######################################################################

Observations of synchrotron intensity and spectral index provide
essential and stringent constraints on the interstellar electron
spectrum and on the magnetic field.

\begin{figure}[!t]%%%%%%%%%%%%%%%%%%%%%%%%%%%%%%%%%%%%%%%%%%%%%%%%%%%%%
\vskip 2.6in
%\special{psfile=diffuse_f7.ps voffset=-110 hoffset=30 vscale=65 hscale=65}
\includegraphics{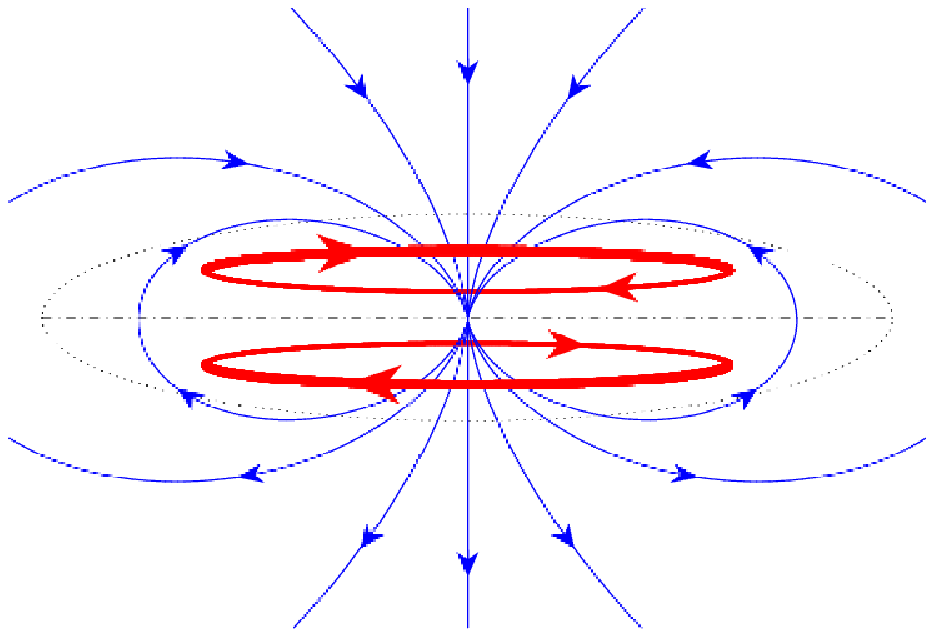}
\caption{The rotation measures of extragalactic radio sources show
the antisymmetric field structure of the Galactic halo (A0 dynamo). 
Adapted from Han (2003). \label{fig:han1}}
\end{figure}

\begin{figure}[!t]%%%%%%%%%%%%%%%%%%%%%%%%%%%%%%%%%%%%%%%%%%%%%%%%%%%%%
\vskip 3.5in
%\special{psfile=diffuse_f8.ps voffset=-125 hoffset=10 vscale=80 hscale=80}
\includegraphics{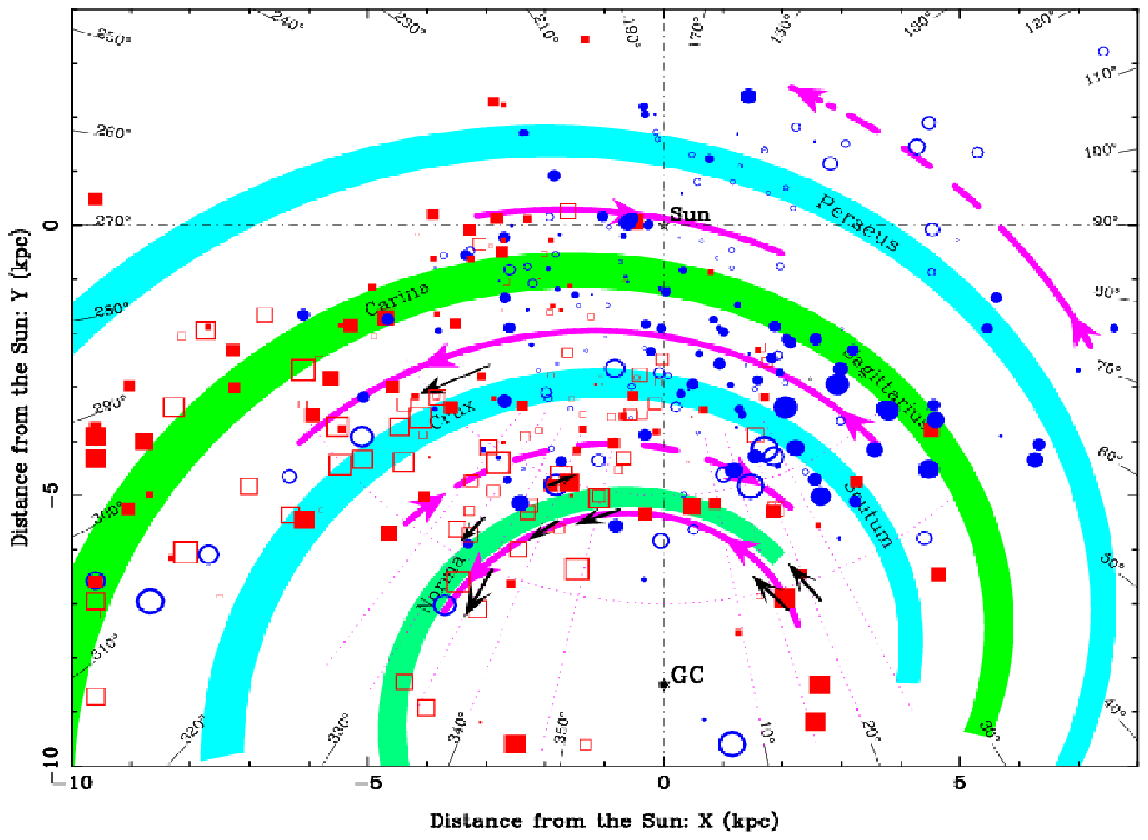}
\caption{The distribution of pulsar rotation measures projected onto 
the Galactic plane reveals the field structure in the Galactic disk,
which has direction reversals from arm to arm.
The well-determined field structure is illustrated by thick lines and arrows.
The thick dashed lines indicate  structures which need further 
confirmation. The symbols are the rotation measures. Adapted from Han (2003). 
\label{fig:han2}}
\end{figure}

The global structure of the Galactic magnetic field is currently derived
from observations of rotation measures of more than 500 pulsars.  It
is best described by two distinct components, (i) a bi-symmetric
spiral field in the disk with reversed direction from arm to arm, and
(ii) an azimuthal field in the halo with reversed directions below and
above the Galactic plane (Figs.\ \ref{fig:han1}, \ref{fig:han2}).

%%%%%%%%%%%%%%%%%%%%%%%%%%%%%%%%%%%%%%%%%%%%%%%%%%%%%%%%%%%

%%%%%%%%%%%%%%%%%%%%%%%%%%%%%%%%%%%%%%%%%%%%%%%%%%%%%%%%%%%

The average strength of the \emph{total} field derived from radio
synchrotron data, under the energy equipartition assumption, is
$6\pm2$ $\mu$G locally and about $10\pm3$ $\mu$G at 3 kpc from the
Galactic center \citep{beck01}.  For comparison, Heiles (1996) gives
$\sim$5 $\mu$G for the volume and azimuthally averaged \emph{total}
field at the solar position. Vall\'ee (1996) gives similar values.
Optical and synchrotron polarization data yield a strength of the
local \emph{regular} field of $4\pm1$ $\mu$G, which is probably an upper
limit. Pulsar rotation measures give a lower value: $1.4\pm0.2$
$\mu$G. The strength of the turbulent magnetic field is $\sim$5 $\mu$G
on typical scale $\sim$50 pc \citep{ohno93}.

%% Double captions:
\begin{figure}[t]%%%%%%%%%%%%%%%%%%%%%%%%%%%%%%%%%%%%%%%%%%%%%%%%%%%%%%
\vskip 2.65in
\includegraphics{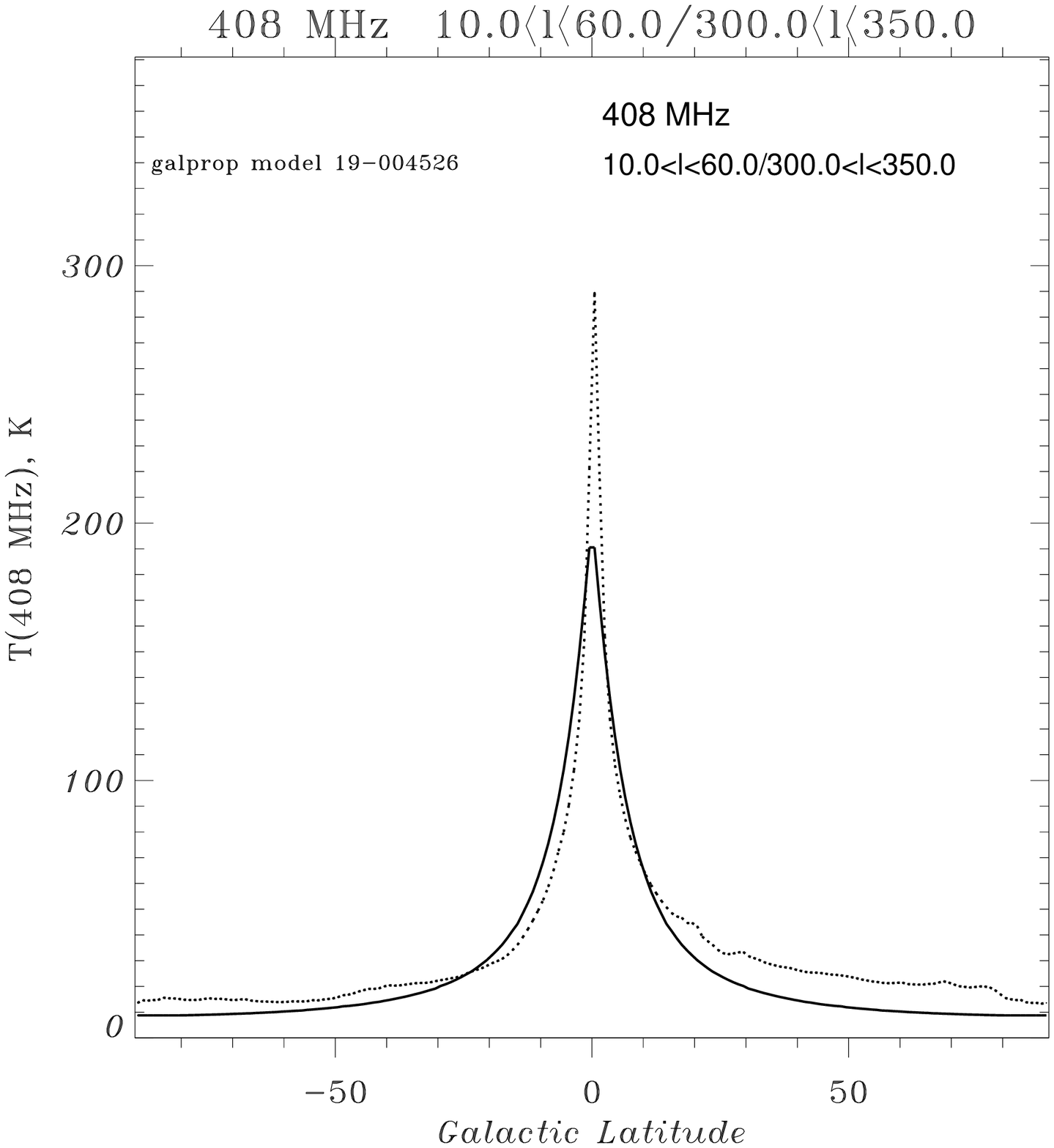}
\includegraphics{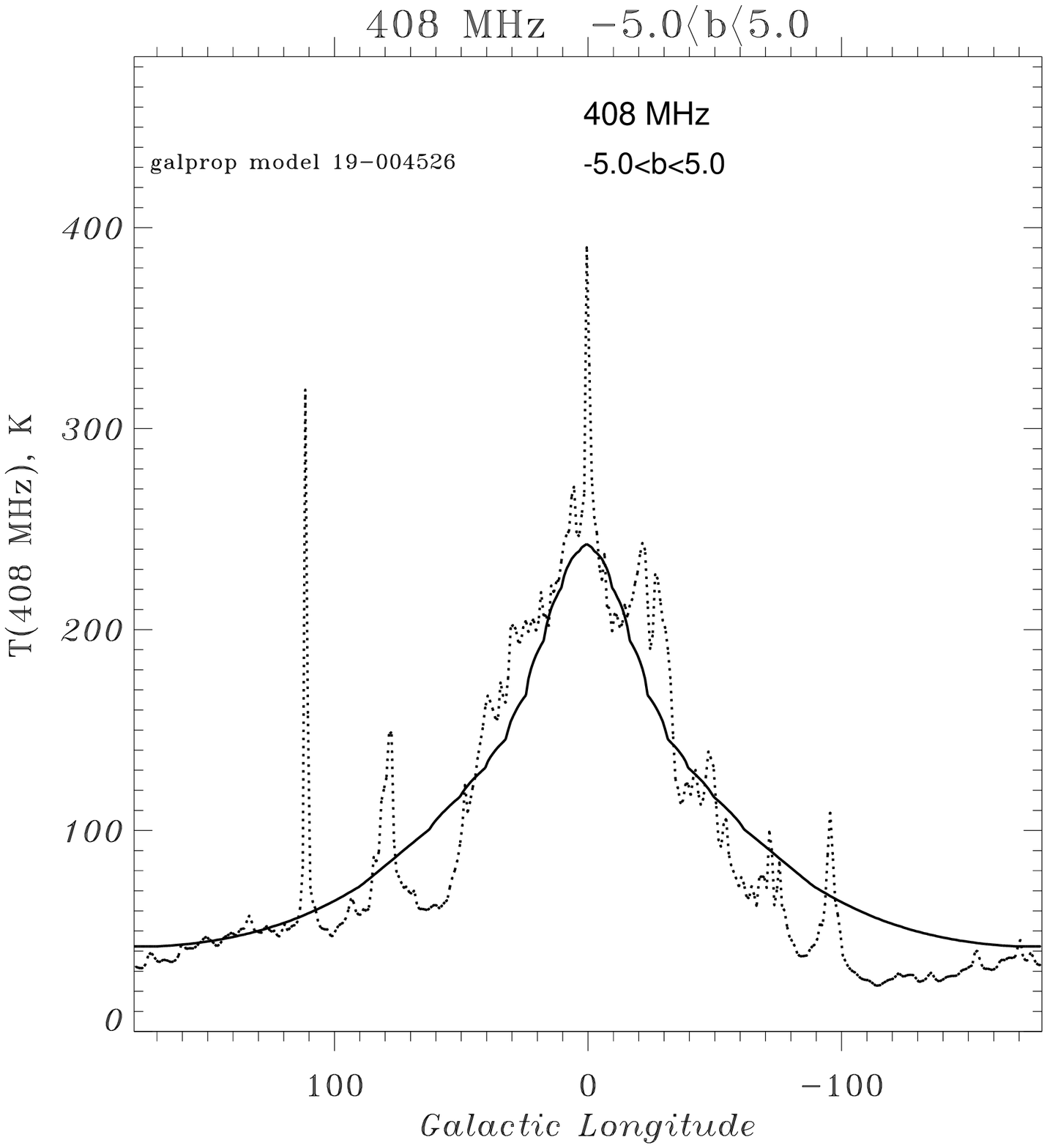}
\caption{Intensity profiles of synchrotron emission at 408 MHz in latitude
($10^\circ\le l\le60^\circ, 300^\circ\le l\le 350^\circ$) 
and longitude as calculated in ``hard electrons and modified nucleons''
(HEMN) model. Data: Haslam et al.\ (1982). Adapted from Strong et al.\ (2000). 
\label{fig:synchrotron}}
\end{figure}%%%%%%%%%%%%%%%%%%%%%%%%%%%%%%%%%%%%%%%%%%%%%%%%%%%%%%%%%%%

The strength of the \emph{total} field thus has a radial scale $R_B = 10$ 
kpc, while a reasonable value for  scale height is $z_B = 2$ kpc,
consistent with radio observations of edge-on spiral galaxies.  Such a
magnetic field reproduces well the absolute magnitude and profiles of
the 408 MHz emission as shown in Fig.\ \ref{fig:synchrotron}.  The
thermal contribution in the plane at this frequency is only about
$\sim$15\% \citep{broadbent89}.   A significantly smaller field would
give too low synchrotron intensites as well as a spectral index
distribution which disagrees with the data.  $R_B$ is constrained by
the longitude profile, and $z_B$ by the latitude profile of
synchrotron emission.  A more detailed fit to the profiles, involving
spiral structure as well as explicit modelling of random and
non-random field components, is given in Phillipps et al.\ (1981),
Broadbent et al.\ (1990), Beuermann et al.\ (1985).

The synchrotron emission in the 10 MHz -- 10 GHz band constrains the
electron spectrum in the $\sim$1--10 GeV range (see e.g.\ Webber et
al.\ 1980).  Out of the plane, free-free absorption is only important
below 10 MHz (e.g., Strong and Wolfendale 1978).  In particular the
synchrotron spectral index ($T\propto\nu^{-\beta}$) provides
information on the ambient electron spectral index $\gamma$ in this
range (approximately given by $\beta = 2 + {\gamma-1\over 2}$).

\begin{figure}[t]%%%%%%%%%%%%%%%%%%%%%%%%%%%%%%%%%%%%%%%%%%%%%%%%%%%%%%
\vskip 1.8in
\includegraphics{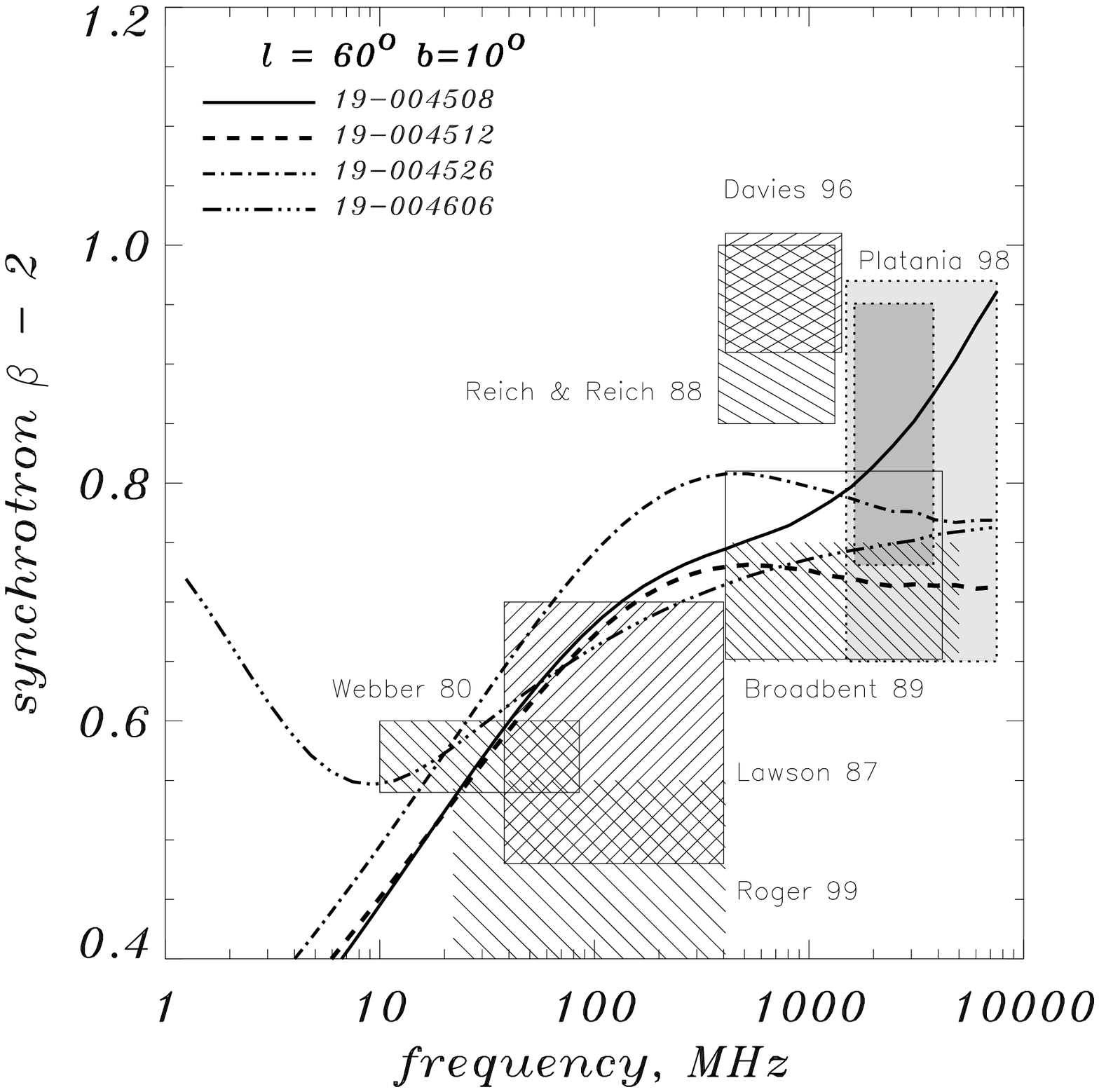}
\narrowcaption{Synchrotron spectral index for selected
propagation models.
Measurements by different authors are shown by boxes.
%Data references are given in the text. 
Adapted from Strong et al.\ (2000).
\label{fig:sych_index}
}
\end{figure}

While there is considerable variation on the sky and scatter in the
observations, and local variations due to loops and spurs, it is
agreed that a general steepening with increasing frequency from
$\beta=2.5$ to $\beta=2.8-3$ is present.  A reanalysis of a DRAO 22
MHz survey \citep{roger99} finds a rather uniform 22 -- 408 MHz
spectral index, with most of the emission falling in the range $\beta
= 2.40-2.55$.  Recent new experiments give reliable spectral indices
up to several GHz \citep{platania98}; they used a catalogue of HII
regions to account for thermal emission.  Fig.\ \ref{fig:sych_index}
summarizes these estimates of the Galactic nonthermal spectral index
as a function of frequency.

%% To make narrow caption:
%%%%%%%%%%%%%%%%%%%%%%%%%%%%%%%%%%%%%%%%%%%%%%%%%%%%%%%%%%%

\section{Diffuse Galactic Gamma-Ray Emission} \label{sec:diffuse}
%######################################################################

The Galactic diffuse continuum \gray\ emission dominates other
components and  has a wide distribution with most emission coming from
the Galactic plane. Its study is important for cosmic ray physics and
lays the ground for other studies  such as extragalactic background
emission.  It is rather easy to get agreement with data within a
factor of $\sim$2 from a few MeV to $\sim$10 GeV with a
``conventional'' set of parameters, but the data quality warrant
considerably better fits.

\begin{figure}[t]%%%%%%%%%%%%%%%%%%%%%%%%%%%%%%%%%%%%%%%%%%%%%%%%%%%%%%
\vskip 0.6in
\includegraphics{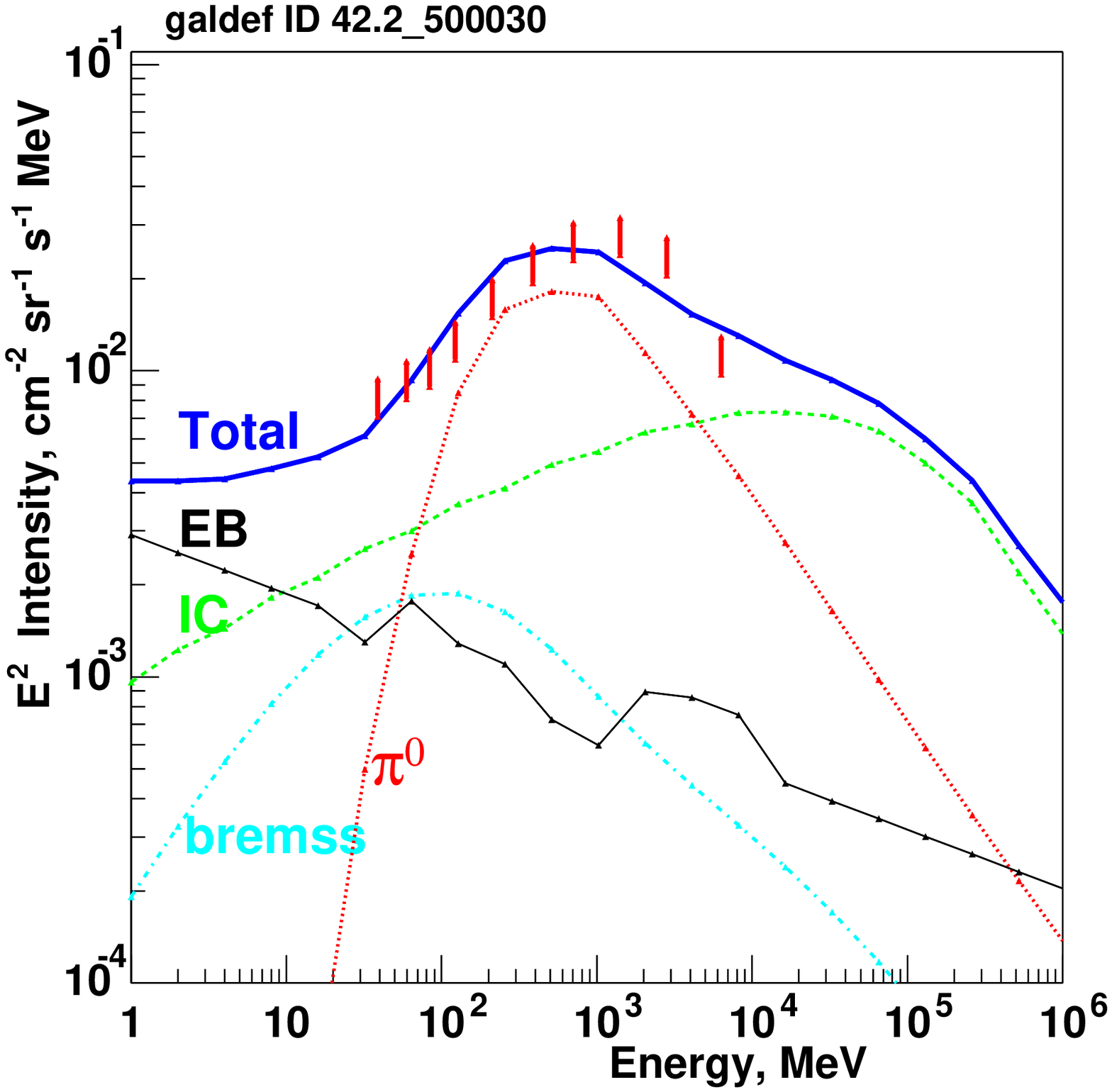}
\narrowcaption{
Spectrum of the Galactic diffuse \gray\
emission from the Galactic plane excluding the inner Galaxy 
($30^\circ<l<330^\circ$, $|b|<5^\circ$). The components
shown are inverse Compton (IC), electron bremsstrahlung (bremss),
$\pi^0$-decay ($\pi^0$), extragalactic diffuse emission (EB).
EGRET data are shown by error bars.
Adapted from Strong et al.\ (2003a). \label{fig:GRB}}
\end{figure}%%%%%%%%%%%%%%%%%%%%%%%%%%%%%%%%%%%%%%%%%%%%%%%%%%%%%%%%%%%

An extensive study of the Galactic diffuse \gray\ emission in the
context of cosmic ray propagation models has been carried out by
Strong et al.\ (2000). This study confirmed that models based on
locally measured electron and nucleon spectra and synchrotron
constraints are consistent with \gray\ measurements in the 30 MeV --
500 MeV range, but outside this range excesses are apparent.  Attempts
were made to explain the observed excess by a harder nucleon spectrum
in the distant regions \cite{mori97,gralewicz97}; however, it seems
that a harder nucleon spectrum is inconsistent with other cosmic ray
measurements such as antiprotons and positrons \cite{moskalenko98}.
The GeV excess appears in all latitude/longitude ranges
\citep{strong03a}.  This implies that the GeV excess is not a feature
restricted to the Galactic ridge or the gas-related emission. A simple
re-scaling of the  components ($\pi^0$, inverse Compton) does not
improve the fit in any region,  since the observed peak is at an energy
higher than the $\pi^0$-peak.  This is an argument towards a
substantial inverse Compton component at high energies.  We note that
a population of unresolved sources can not help to explain the excess
either, since the excess is also present at high Galactic latitudes.

An electron injection index of 1.9 (no breaks) is found optimal,
consistent with earlier findings \citep{strong00} and  observations of
SNRs.  The \emph{average} spectral index of the observed \emph{flux
density} of synchrotron emission from shell type SNRs is close to 0.5
($\beta\sim2.5$),  as expected from Fermi acceleration, implying that
electron spectra there are close to $E^{-2}$ \citep{green01}.  It is
noticeable that small young shell SNRs have steeper spectra, while
older SNRs have generally flatter spectra.
%The filled-center SNRs also have flatter spectra.

In order to be consistent with EGRET data above 10 GeV, a cutoff in
the electron spectrum at 3 TeV is required. The overall quality of the
fit is good.  Fig.\ \ref{fig:GRB} shows the spectrum of the Galactic
diffuse \gray\ emission from the Galactic plane excluding the inner
Galaxy  ($30^\circ<l<330^\circ$, $|b|<5^\circ$).  At low latitudes in
the inner Galaxy the peak around 1 GeV is not  reproduced. To be
consistent at all latitude/longitude ranges, the model required an
adjustment of inverse Compton component via the electron injection
spectrum and a hard spectrum \gray\ compact source population in the inner Galaxy. 
As an
example, the Geminga pulsar does exibit the required hard spectrum
making pulsars a candidate source population.

\begin{figure}[t]%%%%%%%%%%%%%%%%%%%%%%%%%%%%%%%%%%%%%%%%%%%%%%%%%%%%%%
\vskip 2.4in
\includegraphics{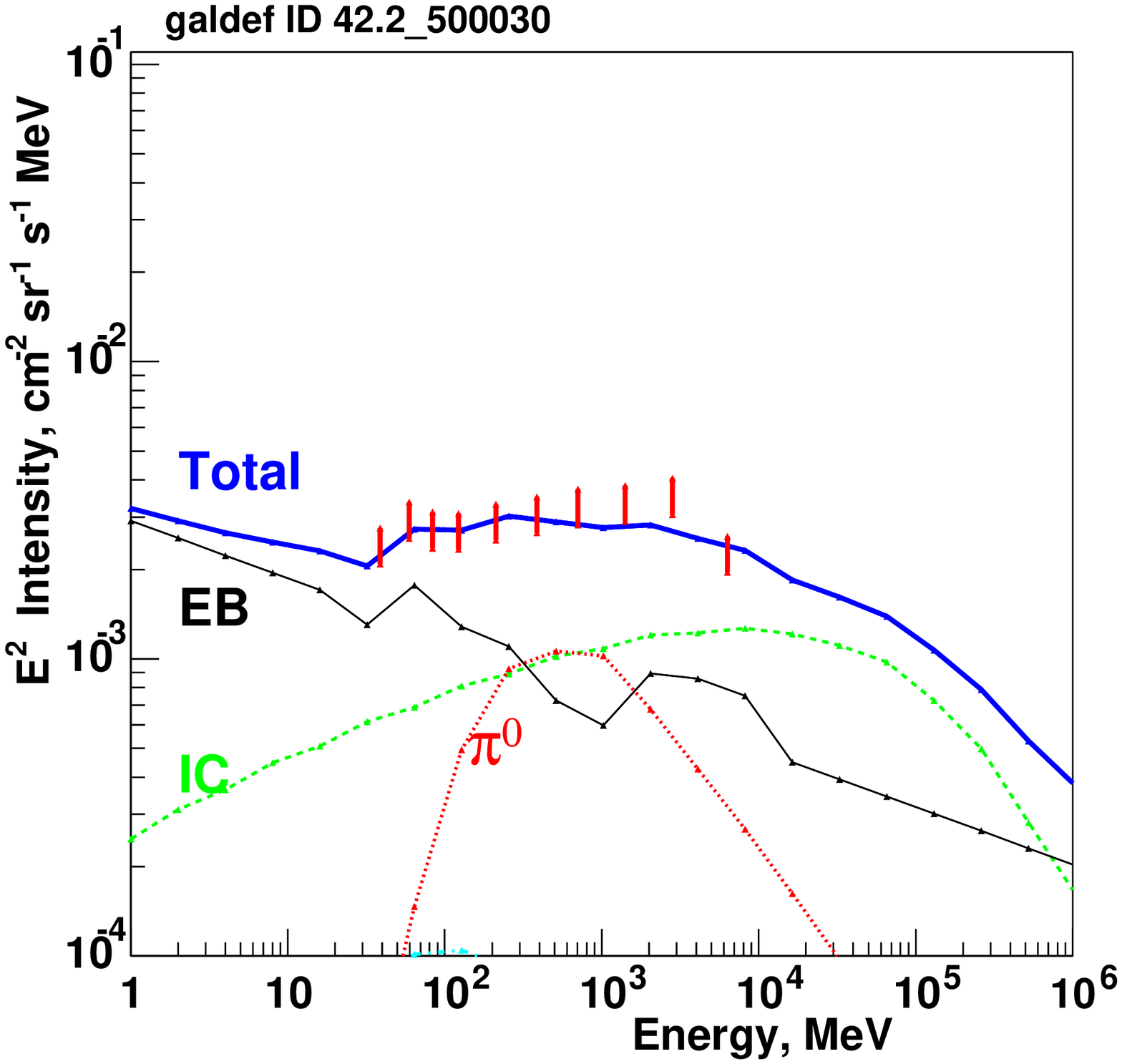}
\includegraphics{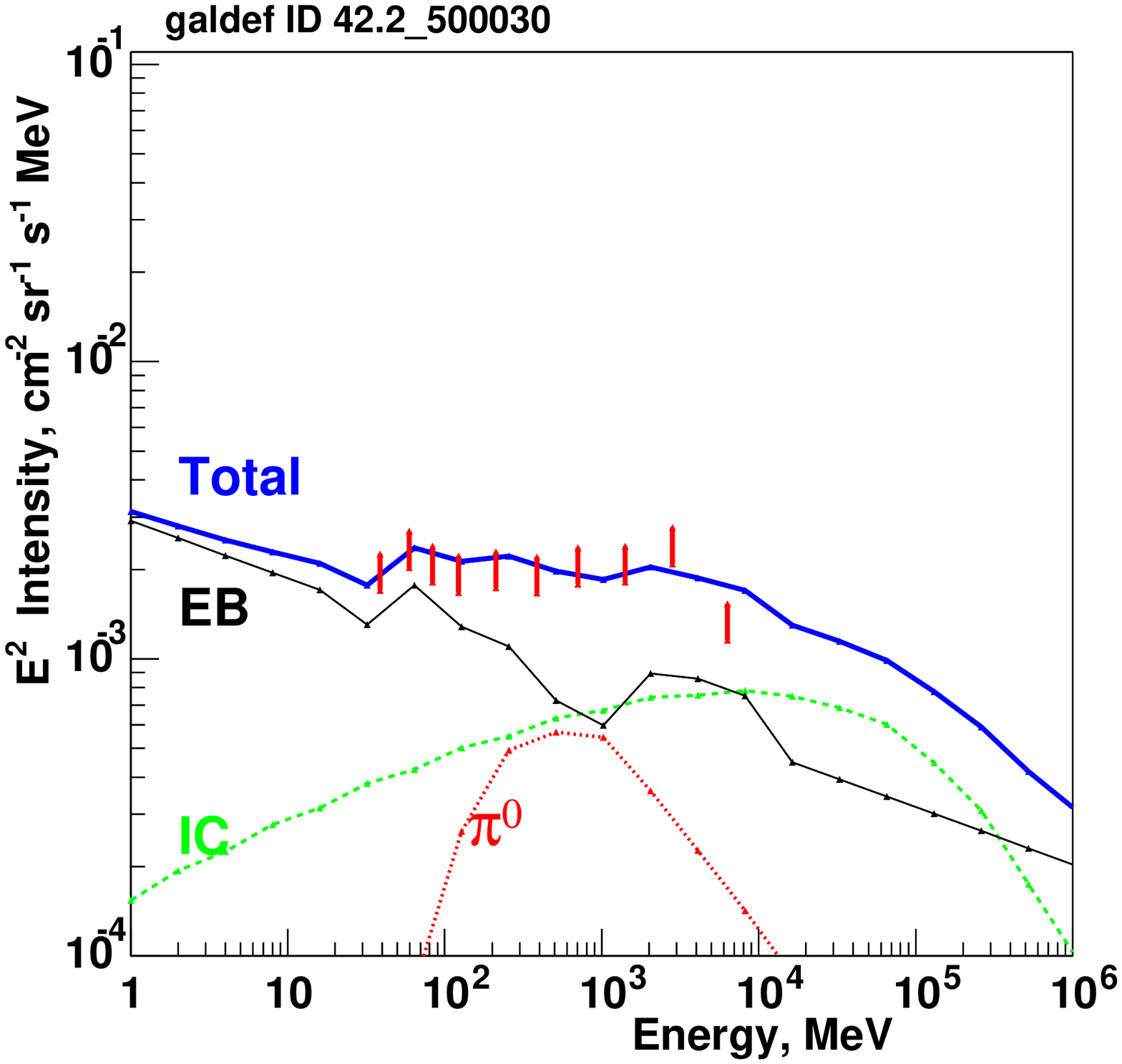}
\caption{
Spectrum of the Galactic diffuse \gray\
emission from high latitudes $0^\circ<l<360^\circ$. 
Left: $20^\circ<b<60^\circ$. Right: $60^\circ<b<90^\circ$.
The lines are coded as in Fig.\ \ref{fig:GRB}.
EGRET data are shown by error bars.
Adapted from Strong et al.\ (2003a). \label{fig:poles}}
\end{figure}%%%%%%%%%%%%%%%%%%%%%%%%%%%%%%%%%%%%%%%%%%%%%%%%%%%%%%%%%%%

The large size of the Galactic halo (4--6 kpc, Moskalenko et al.\ 2001, 
2003) implies that the electron population in the
halo is considerable.  Inverse Compton scattering of photons from the
Galactic plane and CMB  provide a major   contribution to
the Galactic diffuse emission from mid- and high-latitudes.  Fig.\
\ref{fig:poles} shows the energy spectrum of the diffuse emission from
the high Galactic latitudes.  The effect of anisotropic scattering in
the halo \cite{moskalenko00} increases the contribution of Galactic
\grays\ even further and thus reduces the extragalactic component.

\begin{figure}[tp]%%%%%%%%%%%%%%%%%%%%%%%%%%%%%%%%%%%%%%%%%%%%%%%%%%%%%%
\vskip 6.30in
\includegraphics{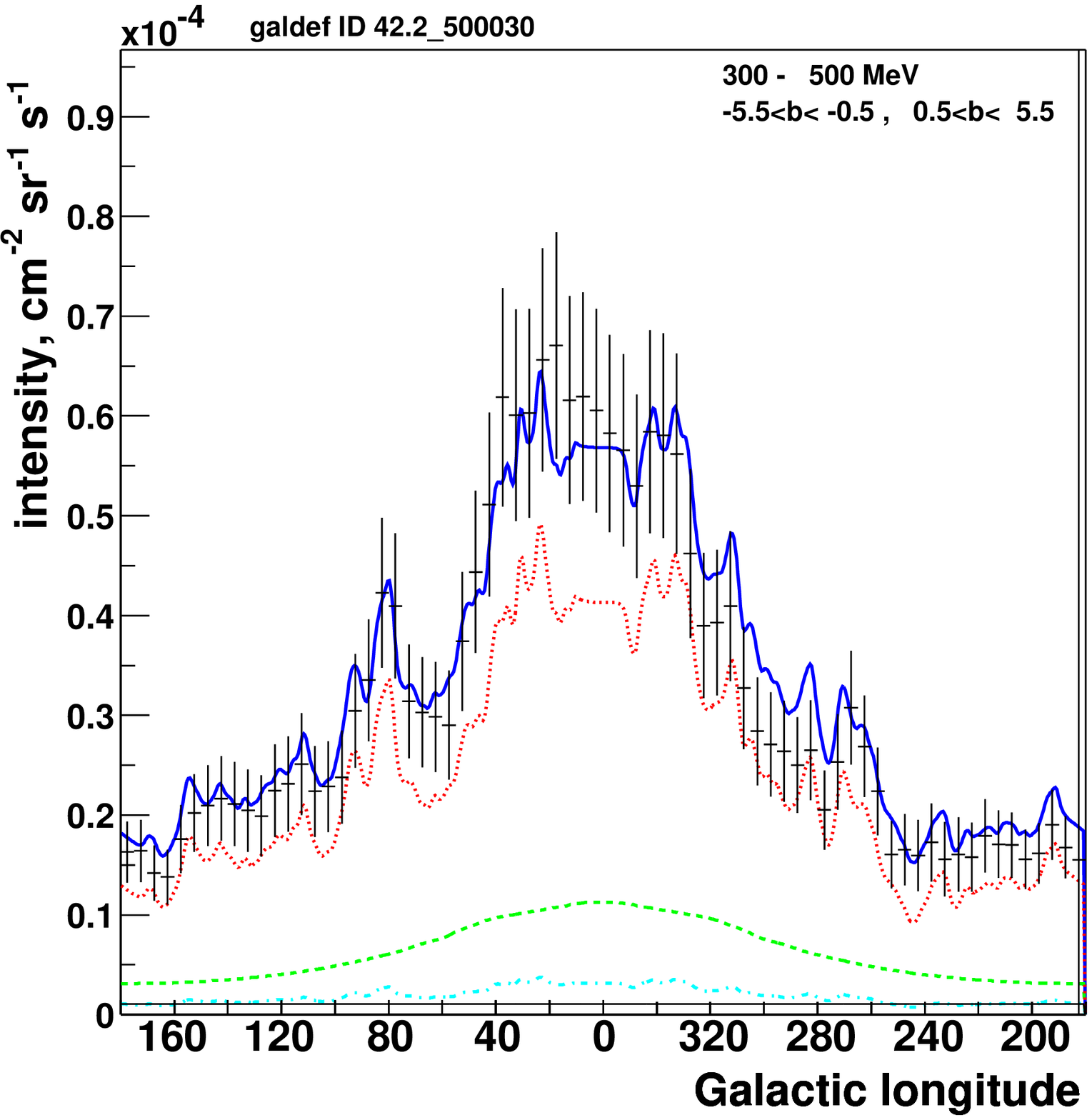}
\includegraphics{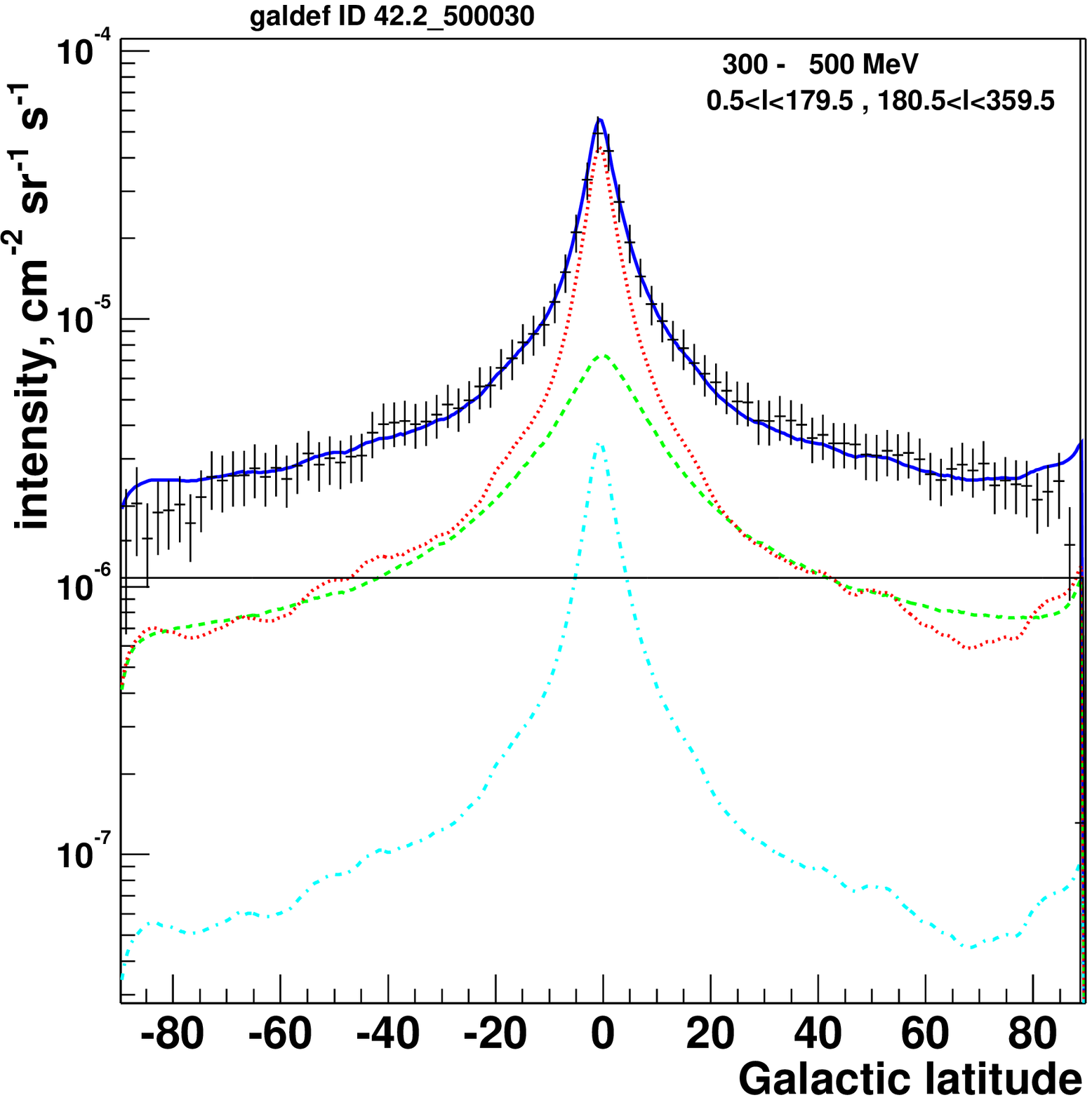}
\caption{
Profiles in longitude along the Galactic plane (top)
and latitude  (bottom) of the Galactic diffuse \gray\
emission in the energy range 300--500 MeV. The components
shown (from top to bottom) are total flux (blue),
$\pi^0$-decay (red), inverse Compton (green), 
electron bremsstrahlung (cyan). The horizontal line is the
extragalactic diffuse emission.
EGRET data are shown by error bars.
Adapted from Strong et al.\ (2004). \label{fig:profiles}}
\end{figure}

Fig.\ \ref{fig:profiles} shows longitude   and latitude profiles of
the diffuse \gray\ emission in the energy range 300--500 MeV.  Because
the Galactic plane is relatively narrow, in latitude the agreement is
always quite good.  In longitude the model appears to be able to
account for  the peaks and dips apparently connected with  details of
the Galactic structure such as spiral arms.

%%%%%%%%%%%%%%%%%%%%%%%%%%%%%%%%%%%%%%%%%%%%%%%%%%%%%%%%%%%

The observations of diffuse TeV emission  from the Galactic plane by
Whipple \cite{lebohec00}, Tibet \cite{amenomori02}, and HEGRA
\cite{aharonian02a} provide only unrestrictive upper limits so far.
A detection of the Galactic plane has been claimed by 
Milagro Collaboration \cite{milagro}.
Interestingly, diffuse TeV \grays\ have been detected from the nearby
(2.5 Mpc) normal spiral starburst galaxy NGC 253  \cite{itoh02}.

\subsection{Analysis of Cosmic Ray Spectral Fluctuations}
%######################################################################

\begin{figure}[t]%%%%%%%%%%%%%%%%%%%%%%%%%%%%%%%%%%%%%%%%%%%%%%%%%%%%%%
\vskip 1.8in
%\special{psfile=diffuse_f14.ps voffset=-95 hoffset=-20 vscale=70 hscale=70}
\includegraphics{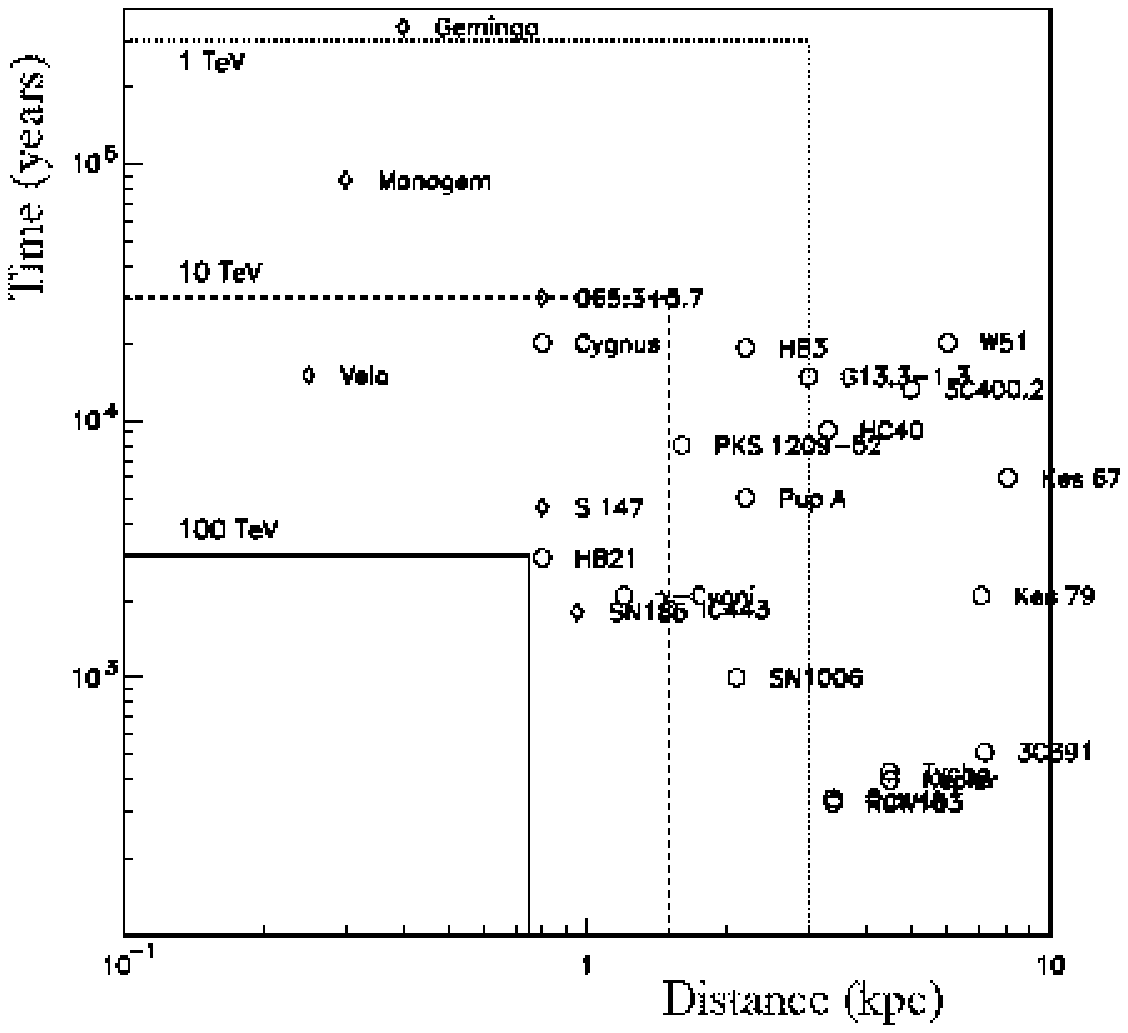}
\narrowcaption{Compilation of nearby shell-type and plerion SNR distances
and ages.  Line boxes represent limiting energies for electrons to
reach Earth \citep{swordy}.
\label{fig:swordy}}
\end{figure}%%%%%%%%%%%%%%%%%%%%%%%%%%%%%%%%%%%%%%%%%%%%%%%%%%%%%%%%%%%

\begin{figure}[t]%%%%%%%%%%%%%%%%%%%%%%%%%%%%%%%%%%%%%%%%%%%%%%%%%%%%%%
\vskip 0.4in
%\special{psfile=diffuse_f15.ps voffset=-160 hoffset=-10 vscale=83 hscale=63}
\includegraphics{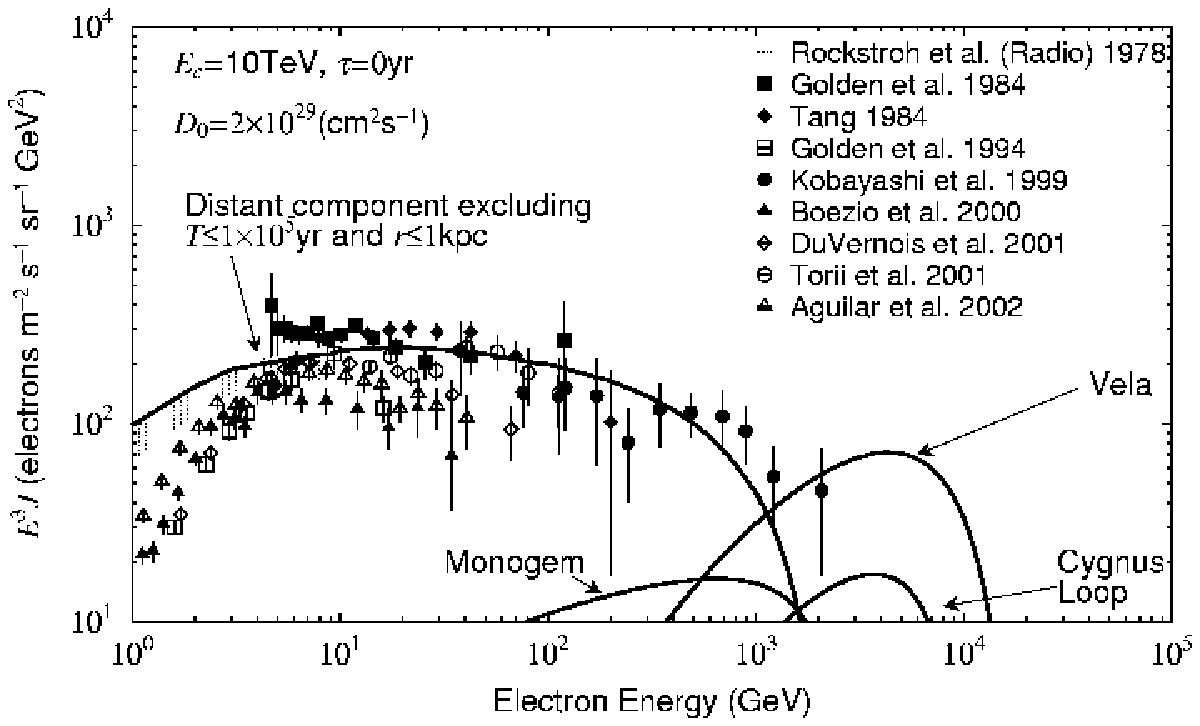}
\narrowcaption{Absolute differential energy spectrum of electrons in 
comparison with calculated results for a diffusion model assuming a power law
index of the injection spectrum 2.4. The contribution of individual sources
is labeled as Vela, Cygnus Loop, and Monogem. Adopted from Kobayashi et al.\ (2003). 
\label{fig:local_electrons}}
\end{figure}

In studies of cosmic-ray propagation and diffuse continuum \gray\
emission from the Galaxy it has usually been assumed that the source
function can be taken as smooth and time-independent. However,
especially for electrons at high energies where energy losses  due to
synchrotron and inverse Compton emission are rapid, the effect of the
stochastic nature of the sources becomes apparent. For the typical
energy density of Galactic radiation and magnetic fields of 1 eV
cm$^{-3}$, the energy loss timescale is $\sim$$3\times10^5$ yr at 1 TeV,
and becomes as short as $\sim$$3\times10^3$ yr at 100 TeV. A cutoff in
the electron spectrum at very high energies is thus unavoidable
because of both large energy losses and a discrete nature of the
sources. This is similar to the GZK effect for ultra high energy
cosmic rays, where the cutoff in the proton spectrum
appears due to the energy losses on photopion production.
The analysis of
nearby shell-type SNRs predicts that the electron spectrum should have
a cut off between 30 TeV and 100 TeV as measured near the solar system
(Fig.~\ref{fig:swordy}). Studies of  the propagation of
very-high-energy electrons from local sources \citep{nishimura97}  has
shown that some nearby SNRs are possibly capable of producing unique
identifiable features in the cosmic-ray electron spectrum at 1--30
TeV, where the important parameters are the distance and the age of a
SNR.  The most promising candidate sources of TeV electrons are Vela,
Cygnus Loop, and Monogem (Fig.~\ref{fig:local_electrons}).
Very-high-energy electron measurements give a direct test of SNR
origin of cosmic rays, but also an important test of our local
environment. The features in the electron spectrum and the cutoff
energy would immediately signal which SNR(s) is/are affecting the
local cosmic-ray flux and to what degree, with implications for
Galactic cosmic-ray propagation models and predictions of the diffuse
\gray\ emission.

%%%%%%%%%%%%%%%%%%%%%%%%%%%%%%%%%%%%%%%%%%%%%%%%%%%%%%%%%%%

\begin{figure}[t]%%%%%%%%%%%%%%%%%%%%%%%%%%%%%%%%%%%%%%%%%%%%%%%%%%%%%%
\vskip 2in
\includegraphics{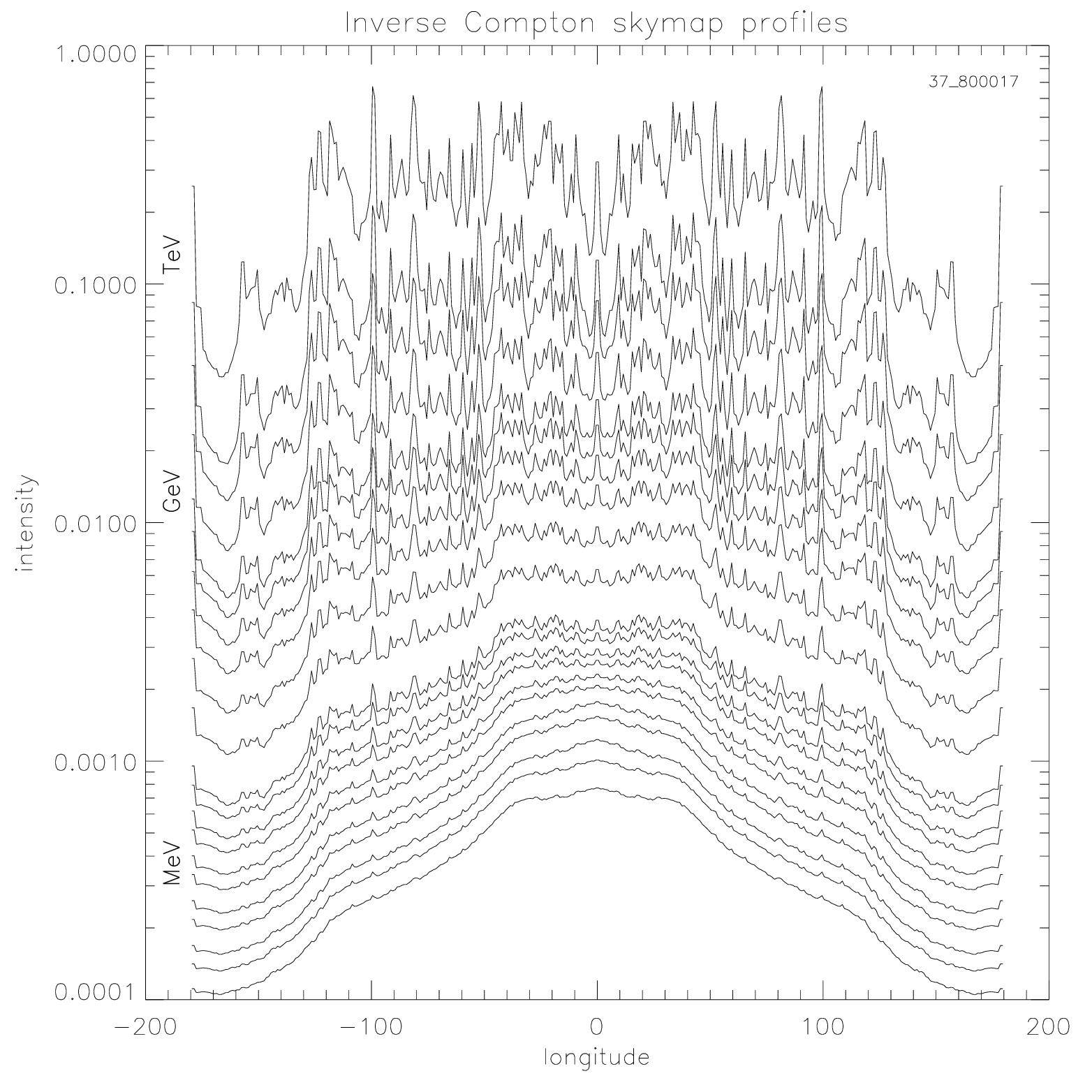}
\narrowcaption{Modeled
inverse Compton \gray\ longitude distributions for \gray\ energies from 1 MeV 
(bottom) to 1 TeV (top). Adopted from Strong and Moskalenko (2001c). 
\label{fig:ics_fluctuations}}
\end{figure}

The fluctuations of electron spectra \emph{for different sources}  has been
invoked to explain the GeV excess in the diffuse emission observed by
EGRET. In particular,  Pohl and Esposito (1998) allowed  the electron
injection index in individual sources to fluctuate  around 2.0, which
would lead to a flatter electron spectrum at high energies and produce
more inverse Compton emission.  In order to include fluctuations in
the source spectra in the cosmic ray propagation code GALPROP, a
model with explicit time-dependence and a stochastic SNR population
has been developed by Strong and Moskalenko (2001a),  which follows
the propagation in three dimensions.  The important parameters here
are the mean time between the events $t_{\rm SNR}$ in a 1 kpc$^3$ unit
volume, and the time of the active phase $t_{\rm cr}$ during which an
SNR produces cosmic rays.  Apparently, the inverse-Compton emission
becomes increasingly clumpy at high energies due to the effect of
individual SNRs as shown in longitude distributions obtained from the
model (Fig.\ \ref{fig:ics_fluctuations}).  The effect is already
visible at 1 GeV and will be an important signature for the GLAST
\gray\ observatory, which will measure up to 300 GeV.

%%%%%%%%%%%%%%%%%%%%%%%%%%%%%%%%%%%%%%%%%%%%%%%%%%%%%%%%%%%

\begin{figure}[t]%%%%%%%%%%%%%%%%%%%%%%%%%%%%%%%%%%%%%%%%%%%%%%%%%%%%%%
\vskip 2.3in
%\special{psfile=diffuse_f17a.ps voffset=-180 hoffset=-75 vscale=48 hscale=48}
%\special{psfile=diffuse_f17b.ps voffset=-180 hoffset=153 vscale=48 hscale=35}
\includegraphics{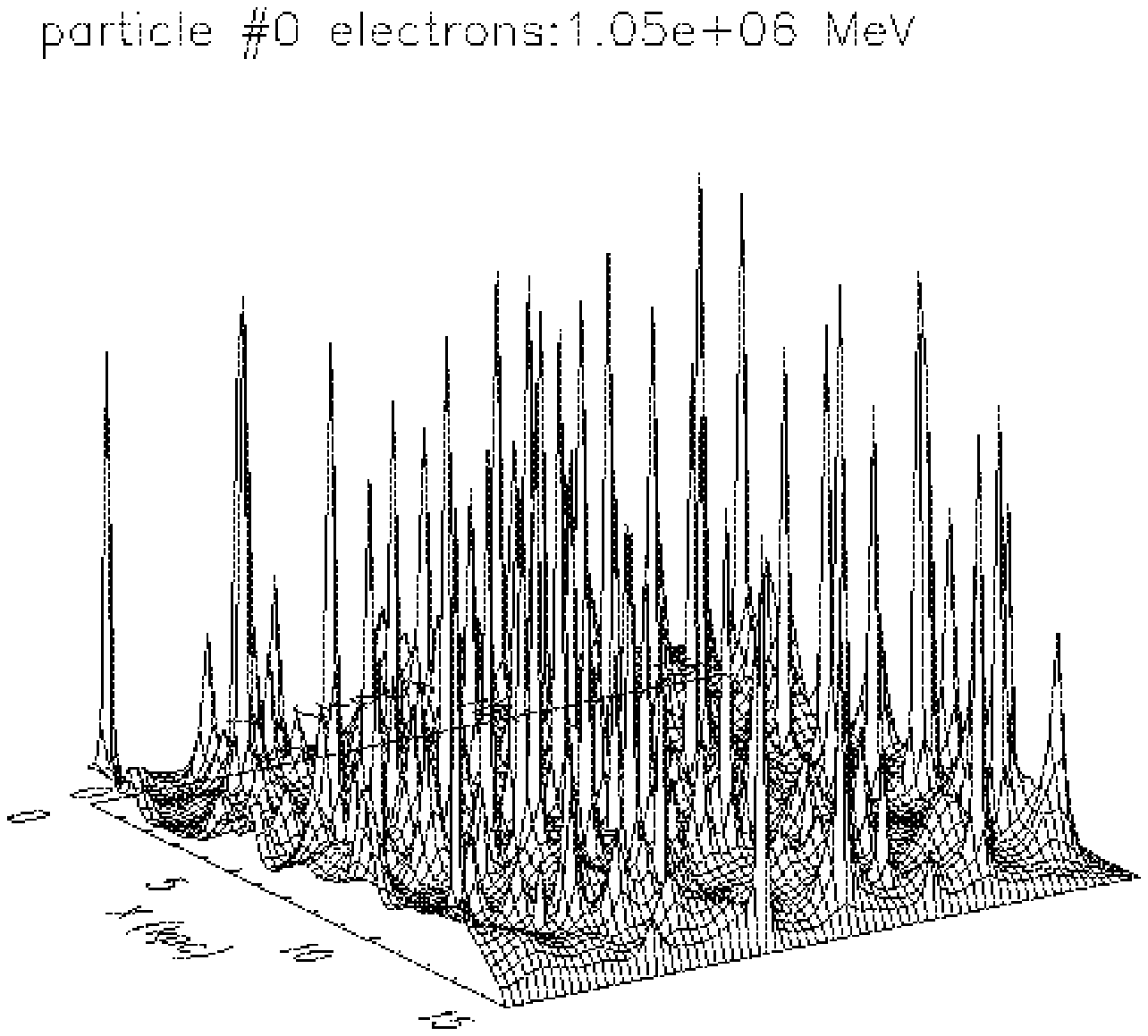}
\includegraphics{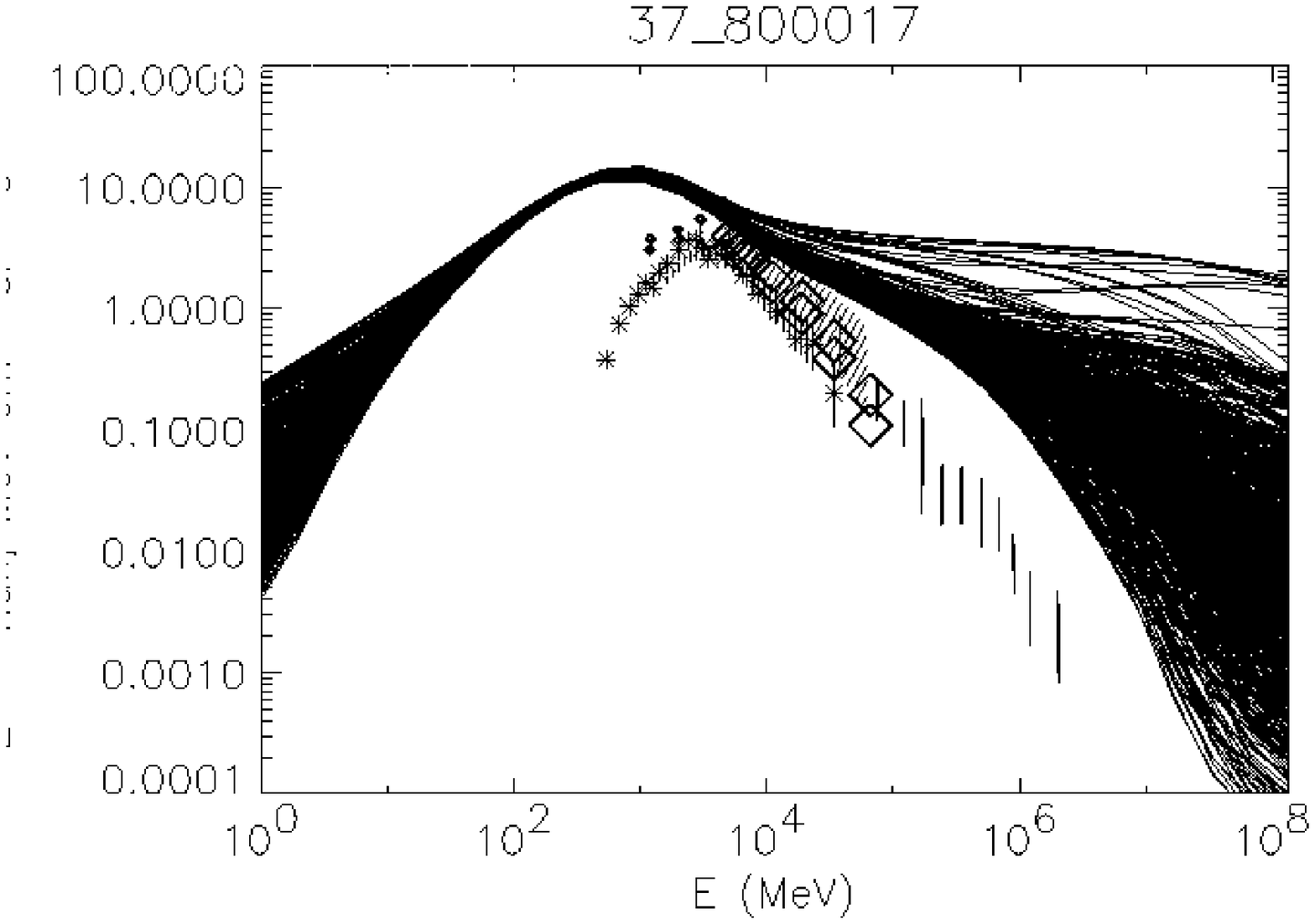}
\caption{Simulated distribution of 1 TeV electrons at $z=0$ (left) and spectral variations 
in $4<R<10$ kpc (right) for $t_{\rm SNR} = 10^4$ yr. 
Data points: locally measured electron spectra. 
Adopted from Strong and Moskalenko (2001a).
\label{fig:elec_fluct_1}}
\end{figure}

\begin{figure}[t]%%%%%%%%%%%%%%%%%%%%%%%%%%%%%%%%%%%%%%%%%%%%%%%%%%%%%%
\vskip 2.3in
%\special{psfile=diffuse_f18a.ps voffset=-180 hoffset=-75 vscale=48 hscale=48}
%\special{psfile=diffuse_f18b.ps voffset=-180 hoffset=153 vscale=48 hscale=35}
\includegraphics{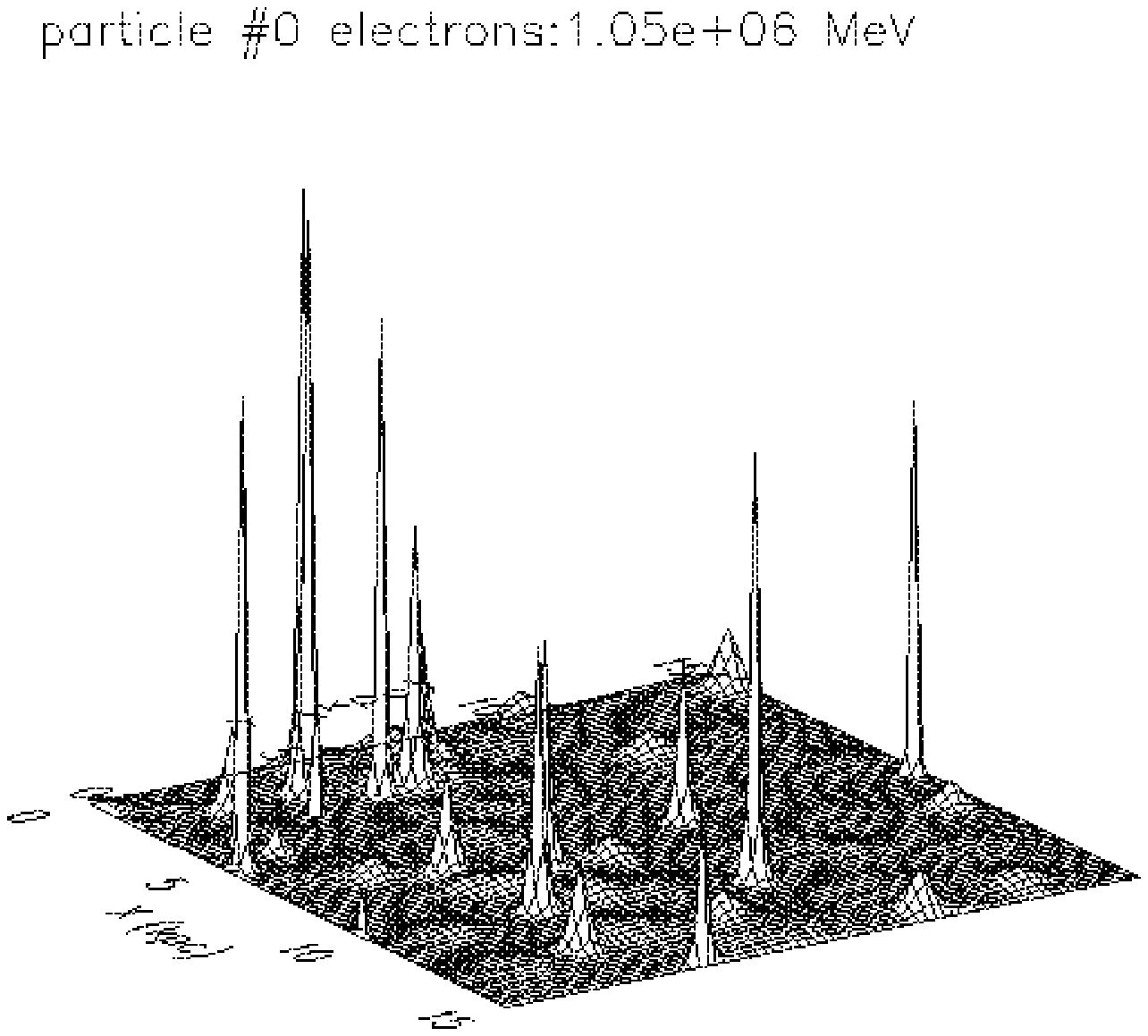}
\includegraphics{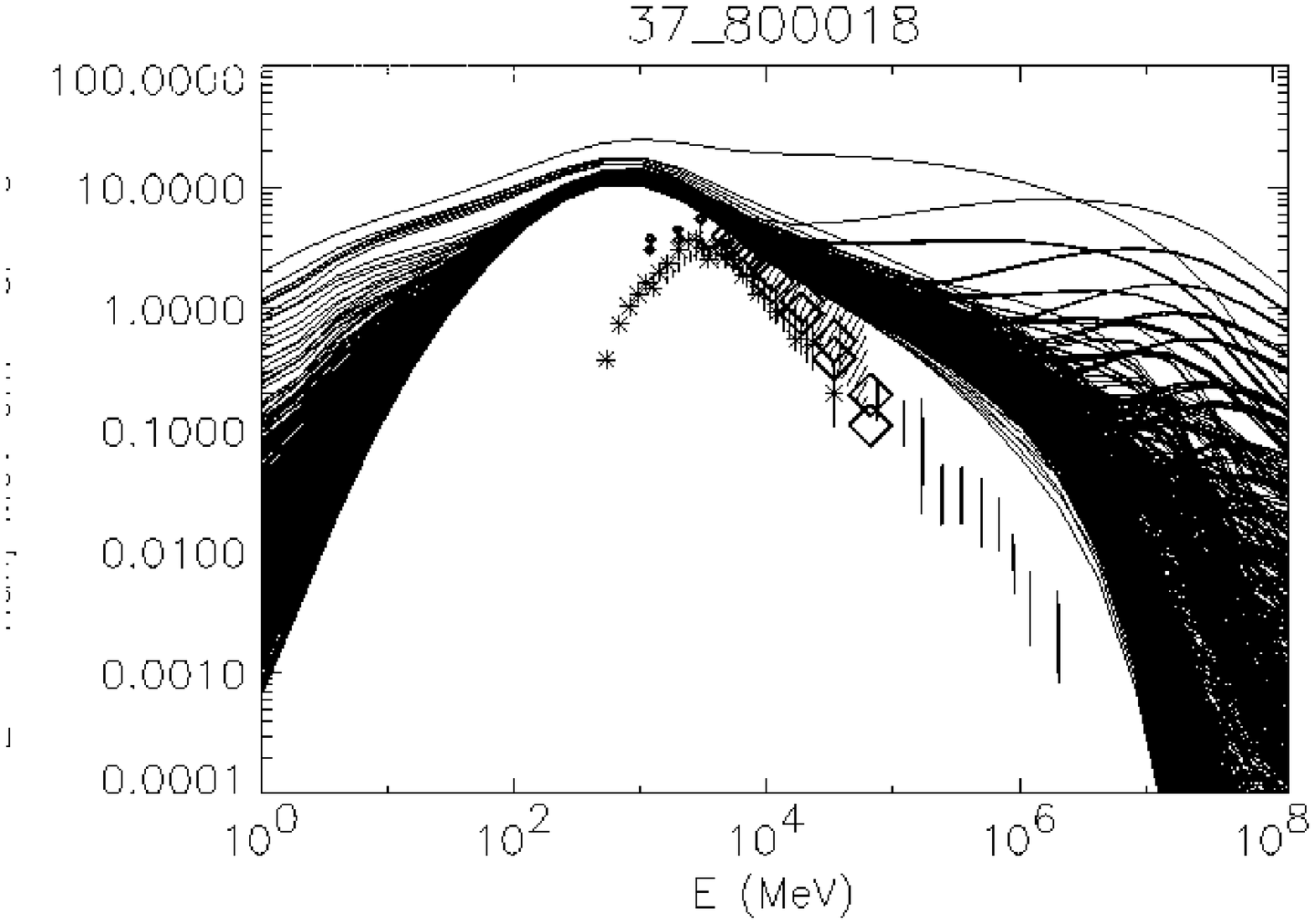}
\caption{Simulated distribution of 1 TeV electrons at $z=0$ (left) and spectral 
variations in $4<R<10$ kpc (right) for $t_{\rm SNR} = 10^5$ yr.
Data points: locally measured electron spectra. 
Adopted from Strong and Moskalenko (2001a).
\label{fig:elec_fluct_2}}
\end{figure}

The results, however, indicate that although the inhomogeneities are
large they are insufficient to easily explain the GeV excess.  Fig.\
\ref{fig:elec_fluct_1} shows the simulated distribution of electrons
of 1 TeV for a ``standard'' Galactic SN rate 3/century ($t_{\rm
SNR}=10^4$ yr). At GeV energies the distribution shows only small
fluctuations, the particle density being dominated by the long storage
times. At higher energies the losses increase and the fluctuations
become significant as the individual SNR events leave their imprint on
the distribution. The TeV electron distribution is quite
inhomogeneous,  but still none of the spectra around $R = R_\odot$
resembles even remotely  that observed locally.  For the Galactic SN
rate 0.3/century ($t_{\rm SNR}=10^5$ yr) the simulated distribution
above 100 GeV is even more inhomogeneous and the spectrum fluctuates
even more (Fig.\ \ref{fig:elec_fluct_2}).   Some of the spectra
resemble that observed locally within a factor of a few, although
still none is fully compatible with the local spectrum.

\begin{figure}[t]%%%%%%%%%%%%%%%%%%%%%%%%%%%%%%%%%%%%%%%%%%%%%%%%%%%%%%
\vskip 2.5in
%\special{psfile=diffuse_f19a.ps voffset=-180 hoffset=-85 vscale=50 hscale=50}
%\special{psfile=diffuse_f19b.ps voffset=-180 hoffset=90 vscale=50 hscale=50}
\includegraphics{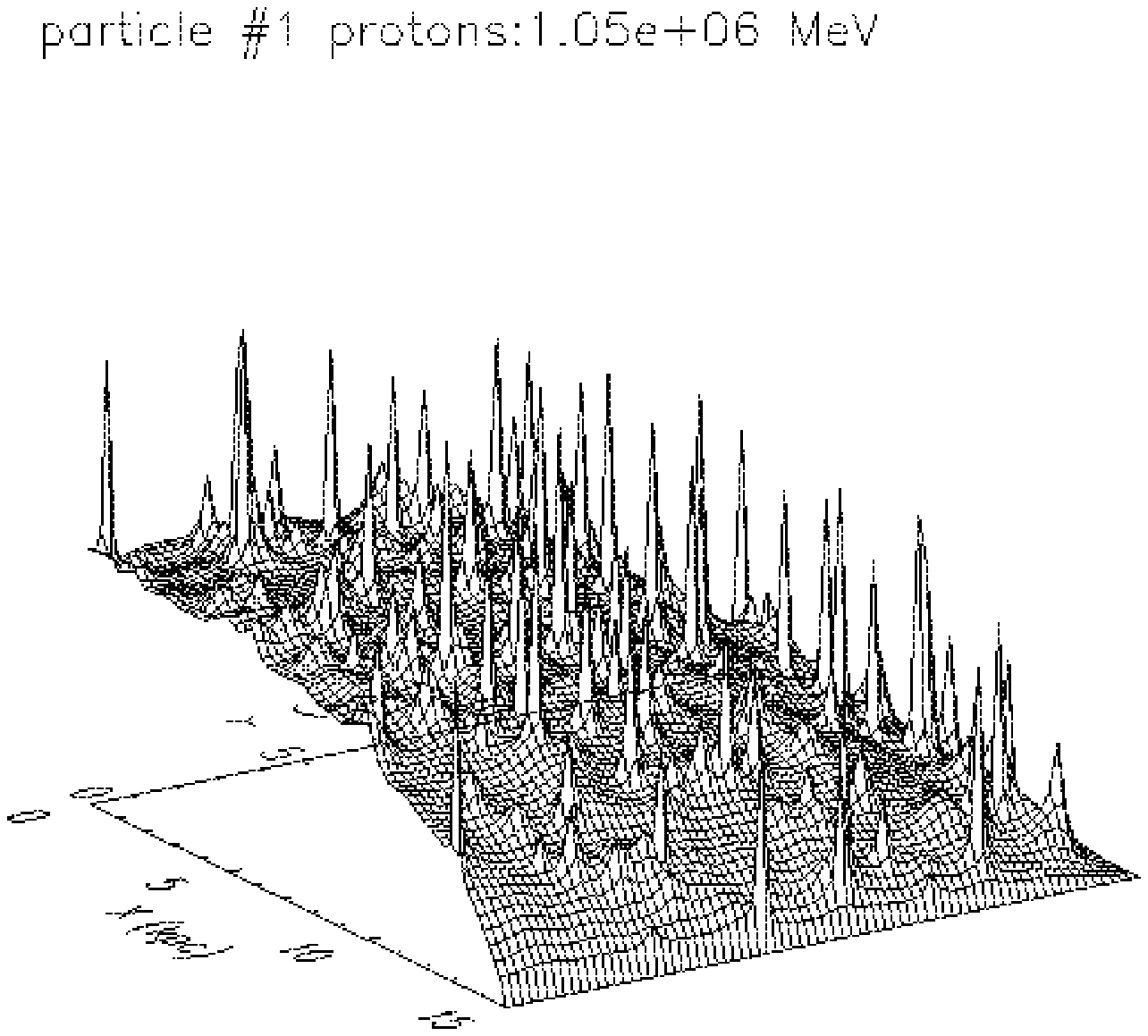}
\includegraphics{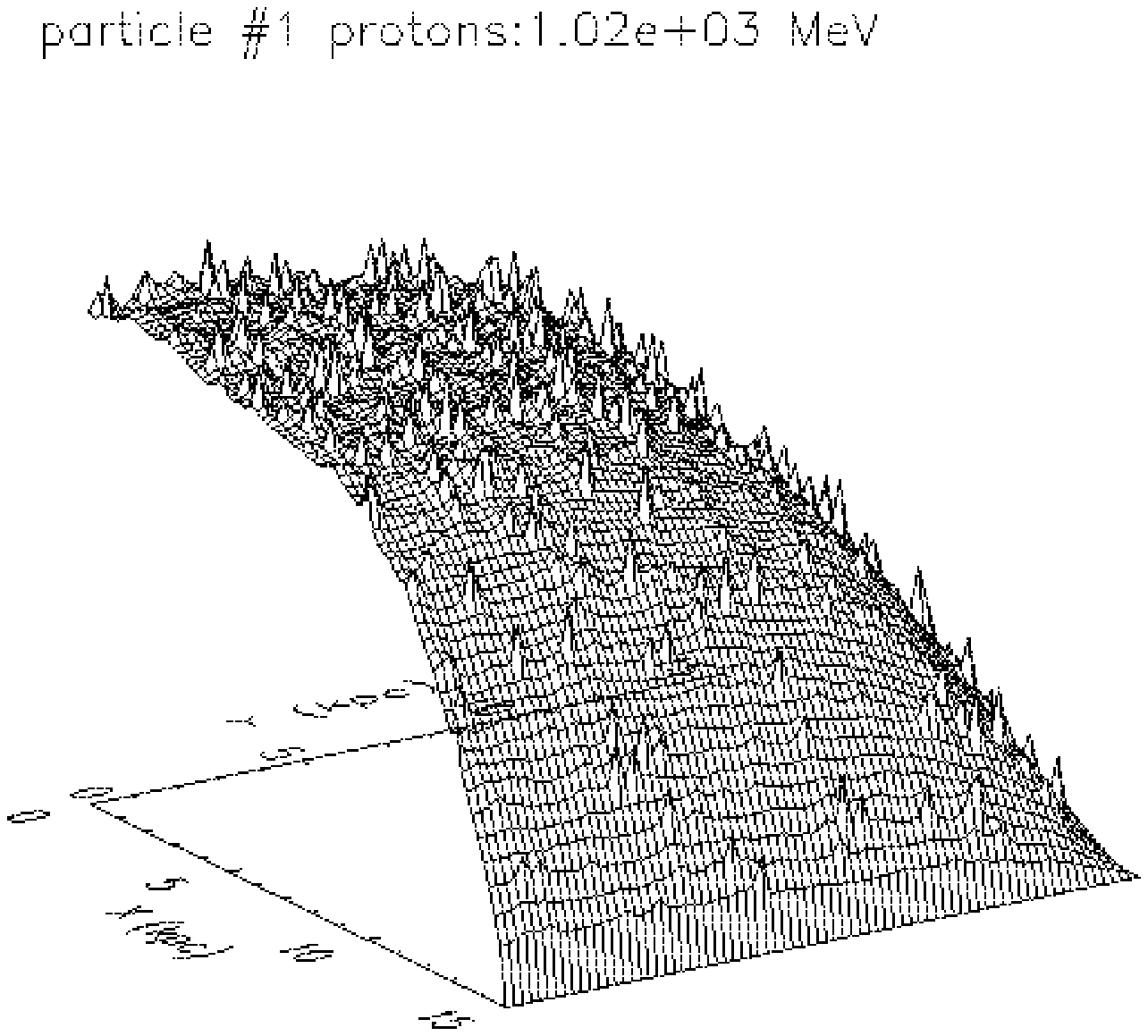}
\caption{Simulated distribution of 1 GeV (left) and 1 TeV (right) protons at $z=0$.
Adopted from Strong and Moskalenko (2001b).
\label{fig:prot_fluct_1}}
\end{figure}

In the case of protons, the fluctuations are also evident,  but much
smaller than for electrons \citep{SM01b}. Fig.\ \ref{fig:prot_fluct_1}
shows the distribution of protons in the Galactic plane ($z=0$) for a
representative quadrant, at two energies.  For illustration we show
results for a model with reacceleration based on Strong et al.\
(2000), and a Galactic SN rate of 3 SN/century.  The stochastic SNR
source produce fluctuations, which are a minimum around 1 GeV and
increase at low energies due to energy losses and at high energies
where the storage of particles in the Galaxy is much reduced so that
the effect of sources  manifests itself on the distribution. Note that
the nature of the fluctuations is different at low and high energies.
However, large fluctuations of the \emph{average} nucleon spectrum are
ruled out on the basis of the ``antiproton test'' proposed by
Moskalenko et al.\ (1998) and confirmed by recent measurements of the
high energy antiproton flux \cite{beach}.

%%%%%%%%%%%%%%%%%%%%%%%%%%%%%%%%%%%%%%%%%%%%%%%%%%%%%%%%%%%

%%%%%%%%%%%%%%%%%%%%%%%%%%%%%%%%%%%%%%%%%%%%%%%%%%%%%%%%%%%

%%%%%%%%%%%%%%%%%%%%%%%%%%%%%%%%%%%%%%%%%%%%%%%%%%%%%%%%%%%

The effect of nearby SNRs on the cosmic-ray anisotropy at 1--1000 TeV
has been studied by Ptuskin et al.\ (2003).  It has been shown that
inclusion of nearby SNRs improves the agreement of the reacceleration
model with the data, while the most important contributions come from
Vela and S 147.  The  very young and close SNR RX J0852.0--4622 (0.2 kpc,
700 yr) would dramatically change the predicted anisotropy, but the
source is probably still in a free expansion stage with accelerating
particles confined inside the remnant.

\subsection{Local Clouds}
%######################################################################

\begin{table}[b]
\caption[Local clouds.]%<-- this version will appear in List of Tables
{Local clouds.\label{table:clouds}}%<-- this version will appear on page
\begin{tabular*}{\textwidth}{@{\extracolsep{\fill}}lrrl}
\sphline
 &
 &
%\it Galactocentric&
 &
$X$, $10^{20}$ cm$^{-2}$\cr

\it Name      &
\it Longitude &
%\it Distance, kpc &
\it Distance &
\it  K$^{-1}$ km$^{-1}$ s\cr
\sphline

Ophiuchus \citep{hunter94}        &
336$^\circ$--\ 10$^\circ$\ &
%8.4 &
125 pc &
$1.1\pm0.2$\cr

Cepheus \citep{digel96}           &
100$^\circ$--130$^\circ$&
%8.7 &
250 pc &
$0.92\pm0.14$\cr

Orion \citep{digel99}             &
195$^\circ$--220$^\circ$& 
%8.9 &
500 pc &
$1.35\pm0.15$\cr

Monoceros \citep{digel01}         &
210$^\circ$--250$^\circ$& 
%9.2 &
830 pc &
$1.64\pm0.31$\cr

Taurus/Perseus \citep{dg01}&
150$^\circ$--185$^\circ$&
%??? &
140/300 pc&
$1.08\pm0.10$\cr
\sphline
\end{tabular*}
\end{table}

\begin{figure}[t]%%%%%%%%%%%%%%%%%%%%%%%%%%%%%%%%%%%%%%%%%%%%%%%%%%%%%%
\vskip 0.2in 
%\special{psfile=diffuse_f20.ps voffset=-310 hoffset=-13 vscale=90 hscale=90} 
\includegraphics{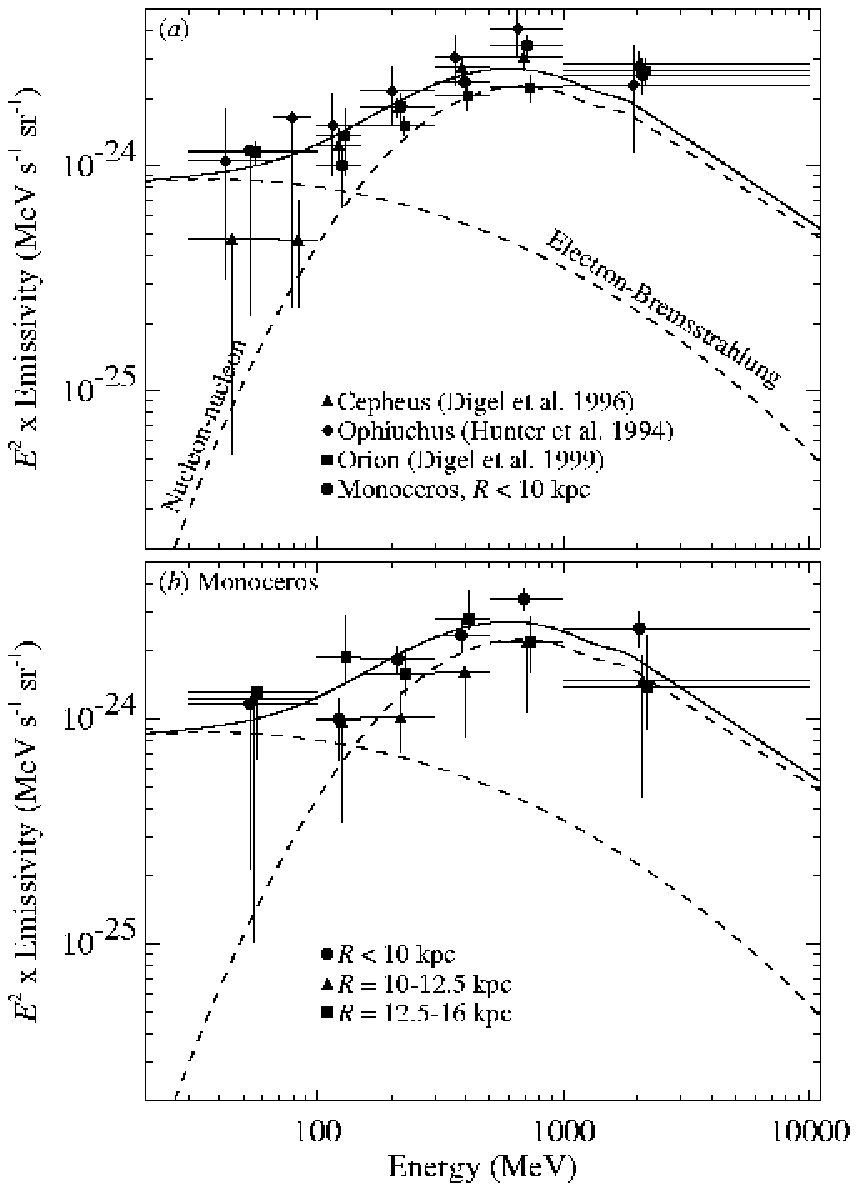} 
\narrowcaption{ (a) Comparison of differential
emissivities for the local gas in Monoceros (circles) with
emissivities from other studies of local clouds: Ophiuchus (diamonds;
Hunter et al.\ 1994), Orion (squares; Digel et al.\ 1999), and Cepheus
(triangles; Digel et al.\ 1996). (b) Differential emissivities in the
three inner annuli in Monoceros. Circles: local ($R<10$ kpc) range;
triangles: interarm ($R=10-12.5$ kpc) range; and squares: Perseus arm
($R=12.5-16$ kpc) range. Also
plotted are the nucleon-nucleon and electron-bremsstrahlung
emissivities and their sum for the solar vicinity. Adopted from Digel
et al.\ (2001). \label{fig:local}}
\end{figure}%%%%%%%%%%%%%%%%%%%%%%%%%%%%%%%%%%%%%%%%%%%%%%%%%%%%%%%%%%%

Some nearby molecular clouds lie at latitudes outside the intense
Galactic plane and hence can be detected as separate extended sources.
The position and distances of clouds observed with EGRET are given in
Table~\ref{table:clouds}.  The \gray\ intensity in these clouds  was
found consistent with that found for the solar circle in large-scale
studies of diffuse emission (Fig.\ \ref{fig:local}).  The differential
\gray\ emissivity is consistent with electron and proton cosmic ray
spectra approximately the same as in the solar vicinity.  This
suggests that the density of cosmic ray protons does not vary
significantly on scales $\lsssim$1 kpc. Interestingly, the Cepheus,
Orion, and Monoceros clouds exibit a ``GeV excess'' similar to that
found in the Galactic plane.

\section{Extragalactic Diffuse Emission} \label{egb}
%######################################################################

The extragalactic diffuse \gray\ background emission (EGB) is the
component of the diffuse emission which is  most difficult to
determine. Its spectrum depends much on the adopted model of the
Galactic background which itself is  not yet firmly established.  It
is not correct to assume that the isotropic component is wholly
extragalactic, because even at the Galactic poles  it is comparable to
the Galactic contribution from  inverse Compton scattering of the
Galactic plane photons   and CMB.  The size of
the halo, the electron spectrum there, and the spectrum of low-energy
background photons are all model dependent and must be derived from  many 
different kinds of observations.

Potentially, if reliably derived, the EGB can provide very important
information about the phase of baryon-antibaryon annihilation
\cite{gao,dolgov}, evaporation of primordial black holes
\cite{hawking,maki}, annihilation of so-called weakly interacting
massive particles (WIMPs) \cite{jkg}, extragalactic IR and optical
photon spectra \cite{stecker}, and/or unresolved sources (AGNs?) and
their cosmological evolution.

Extensive work has been done \cite{sreekumar98} to derive the spectrum
of the EGB based on EGRET data.  The relation of
modelled-Galactic-diffuse-emission vs.\  total-diffuse-emission was
used to determine the EGB as the extrapolation to zero Galactic
contribution.  The derived index $-2.10\pm0.03$ appears to be close to
that of \gray\ blazars.

\begin{figure}[t]%%%%%%%%%%%%%%%%%%%%%%%%%%%%%%%%%%%%%%%%%%%%%%%%%%%%%%
\vskip 3.5in
%\special{psfile=egb.ps voffset=-30 hoffset=400 vscale=55 hscale=55 angle=90}
%\special{psfile=diffuse_f21.ps voffset=-245 hoffset=-195 vscale=46.2 hscale=46.2 angle=0}
\includegraphics{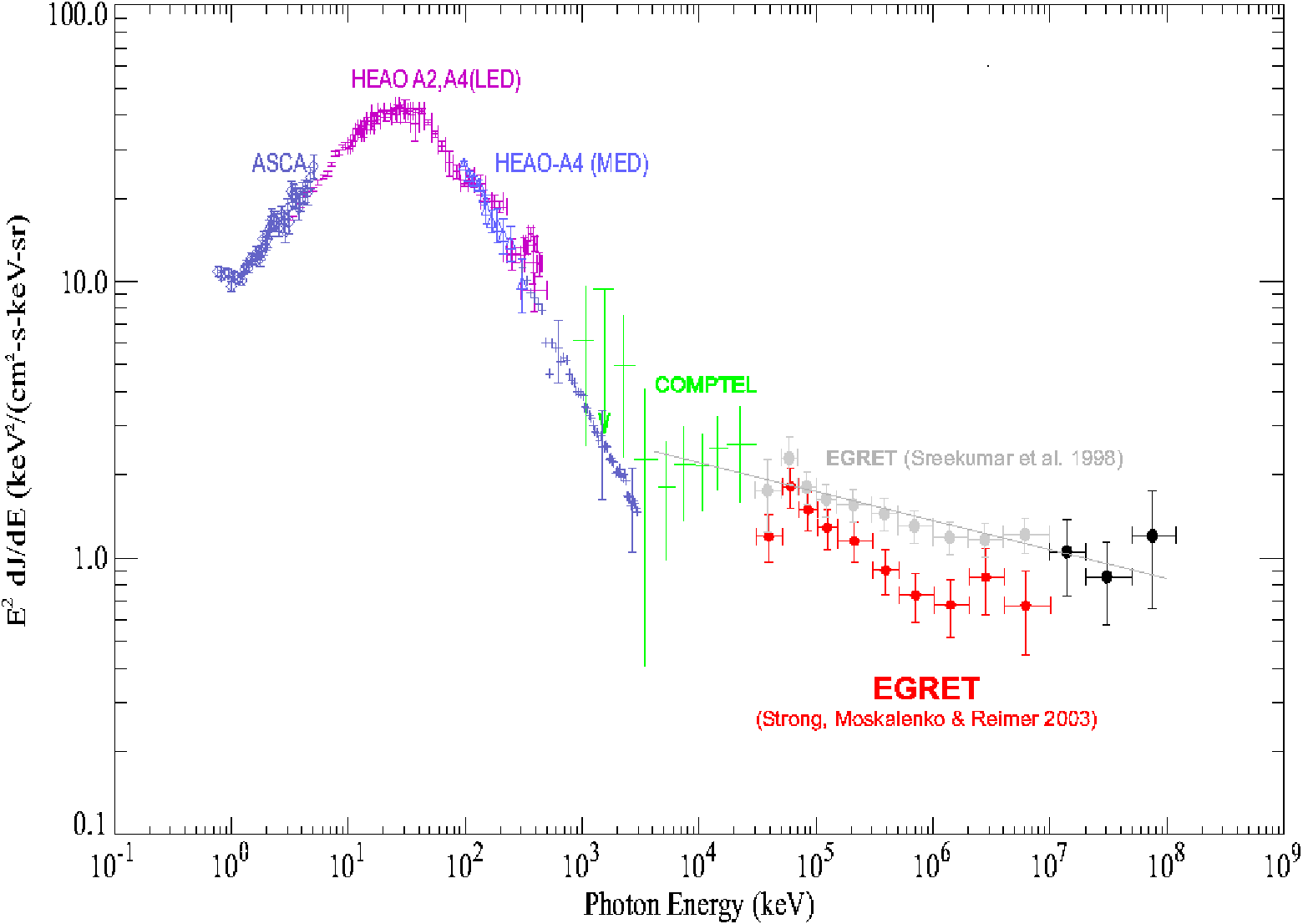}
\caption{ Spectrum of the extragalactic diffuse \gray\
emission.  The solid line (a power-law fit with index --2.1) and the
data points (gray) along the line in the EGRET energy range is the EGB as
derived by Sreekumar et al.\ (1998).  The small circles (red) with error
bars below the line show our new determination of the EGB.  Adapted from
Strong et al.\ (2003b). \label{fig:EXGRB}}
\end{figure}

A new approach to the determination of the EGB is based on cosmic ray
propagation model \cite{strong03a}.  This model reproduces
successfully diffuse \gray\ emission from the entire sky (see previous
Sections.)  To reduce the effects of Galactic structure the fits are
made excluding the plane ($b>10^\circ$). The model gives a good linear
prediction for observed vs.\ predicted \gray\ intensities.  The
spectrum derived appears to be steeper than $-2.10$ and is a smooth
continuation of the extragalactic spectrum at lower energies (Fig.\
\ref{fig:EXGRB}).  There is an indication of a possible upturn at
$\sim$10 GeV.  The positive curvature in the newly determined EGB is
interesting, and is to be expected in the ``unresolved blasar origin hypothesis''
of the EGB \citep{salamon98}.

%%%%%%%%%%%%%%%%%%%%%%%%%%%%%%%%%%%%%%%%%%%%%%%%%%%%%%%%%%%

\section{Faint Sources}
%######################################################################

From the known populations of high-energy \gray\ sources the
contribution from faint sources can be deduced. Faint sources
discussed here include sources below the detection threshold as chosen
during source catalog compilations as well as unresolved sources in
the \gray\ sky.

Clearly, most directly accessible is the dominant class of \gray\
emitters,  the Active Galactic Nuclei (AGN). Known to emit up to the
highest energies, a significant number of not-yet-discovered or
unresolved AGN is expected to contribute to the \gray\ sky. Depending
on the luminosity function of the detected AGN, considerations of high
activity vs.\ low activity states, and the applicability of a blazar
classification/unification scheme quantitative assessments of the
contribution of AGN to the extragalactic background have been made
\cite{stecker96,mucke00,mukherjee,chiang}.  There is a consensus that
blazars should contribute significantly to the observed extragalactic
diffuse emission, however the predictions range from 25\% up to 100\%.
Also, contributions from other extragalactic sources have been
suggested: galaxy clusters might contribute to the extragalactic
\gray\ background either as point-like or extended sources below our
current instrumental detectability, distant \gray\ burst events, or 
a result of large scale cosmological structure formation.

Faint sources will likely contribute also to the diffuse Galactic
emission.  The inner Galactic ridge  is known to be an intense source
of diffuse continuum hard X- and soft \gray\ emission. The hard X-ray
emission was discovered in 1972 \citep{bleach72} and has subsequently
been observed from keV to MeV energies by ASCA, Ginga, RXTE, OSSE,
COMPTEL, Chandra, and most recently by INTEGRAL \citep{strong}.  While
the physical process ($e^+e^-$ annihilation) producing the positron
line and positronium continuum is clear, the source of the remaining
continuum is not, although nonthermal bremsstrahlung is most likely
\citep{dogiel02a}.  The implied photon luminosity of a few 10$^{38}$
erg s$^{-1}$ is remarkable \citep{dogiel02b}. An origin in a
point-source population seems unlikely \citep{tanaka99,tanaka02} since
there are no known candidate objects, and high-resolution imaging with
Chandra shows a truly diffuse component \citep{ebisawa01}.  An
analysis of RXTE data in the inner Galaxy \citep{revnivtsev} indicates
that, after accounting for detected sources, only 10\% of the plane
emission in the 3--20 keV band can be attributed to undetected faint
point sources,  the rest being diffuse.

In the 1--30 MeV range, it appears difficult to account for all the
emission observed by COMPTEL in terms of interstellar processes
(bremsstrahlung/inverse Compton), and hence a significant source
contribution has been proposed \citep{strong00}. There is no
prediction available for how the classes represented by unidentified \gray\
sources might contribute to the observed Galactic diffuse emission.
The contribution of pulsars to the Galactic diffuse emission is
supposedly very little at MeV-energies, but might be in the order of
20\% at GeV's \cite{pohl97}. Only a few pulsars have been 
detected in \grays, and the nature of the majority of Galactic \gray\
sources is still unknown.

\section{Tracers of Exotic Physics ?}
%######################################################################

The nature and properties of the dark matter that may constitute a
significant fraction of the mass of the universe have puzzled
scientists for more than a decade.  Among the favoured dark matter
candidates are WIMPs, whose existence follows from supersymmetric
models.  In most models these particles are stable,  electrically
neutral, lightest neutralino $\chi^0$, which has appropriate 
annihilation cross section and mass to provide suitable relic density.

A number of methods have been proposed to search for evidence for such
particles. These include direct searches for scattering off a nucleus
in a detector, indirect searches to detect the annihilation products,
and collider experiments (for a review and references, see Bergstr\"om
2000).  The indirect searches \cite{jkg} include antiprotons and
positrons in cosmic rays, \grays\ from the Galactic center and halo
(diffuse emission), and neutrinos from massive bodies like the
Galactic center, the sun, and the earth.

In \grays\ the signal could be a relatively narrow line with energy
far beyond that of ordinary particles, or a broad feature appearing as
a result of a decay chain.  The current accelerator limit is
$m_\chi\gtrsim50$ GeV \cite{ellis00}.  GLAST observations will be able
to provide a ``smoking gun'' or to put new limits on supersymmetric
models.

The GeV excess in the EGRET data relative to that expected is
intensively discussed in the literature (see Section \ref{connection}).
Is it a key to the problems
of cosmic-ray physics, a signature of exotic physics (e.g., WIMPs
annihilation, primordial black hole evaporation), or just a flaw in
the current models? This also has an immediate impact on the
extragalactic background radiation studies since its spectrum and
interpretation are model dependent.

Because of the complicated input (see Sections \ref{connection}, 
\ref{sec:diffuse}), the excess can be the result of
incomplete knowledge of the source distribution, the injection
spectra of primary species, the production mechanisms of secondaries,
the interstellar radiation field, or a
combination of these. Some part of the excess can be associated with
cosmic ray sources where freshly accelerated particles interact
with nearby gas particles, producing harder \gray\ spectra.  Therefore,
further deep study of cosmic-ray propagation in a detailed model is
necessary.  Re-evaluation of the interstellar radiation field and the
gas distribution including details of Galactic structure (e.g., spiral
arms) are desirable.  The goal is to develop a model which is
consistent with cosmic-ray data and simultaneously with diffuse \gray\
data \emph{or} clearly indicate the reason for the discrepancy.

\section{Broader Picture and Future Perspective}
%######################################################################

Astrophysics of cosmic rays and \grays\ depends very much on the
quality of the data, which become increasingly accurate each year and
therefore more constraining. While  direct measurements of cosmic
rays are possible in only one location on the outskirts of the Milky
Way, the Galactic diffuse $\gamma$-ray emission provides insights into
the spectra of cosmic rays in distant locations, therefore
complementing the local cosmic-ray studies. This connection, however,
requires extensive modeling and is yet to be explored in detail. The GLAST
mission, which is scheduled for launch in 2007 and is capable of
measuring $\gamma$-rays in the range 20 MeV -- 300 GeV, will change
the status quo dramatically.  The detailed spectra and skymaps of the
Galactic diffuse \gray\  emission gathered by GLAST will require
adequate theoretical models. The efforts will be rewarded by the
wealth of information on cosmic ray spectra and fluxes in remote
locations.  In its turn, a detailed cosmic ray propagation model will
provide a reliable basis for other studies such as search for dark
matter signals in cosmic rays and diffuse \grays, spectrum and origin
of the extragalactic \gray\ emission, theories of nucleosynthesis and
evolution of elements \cite{fields01} etc.  
In addition, GLAST will be able to detect
\grays\ from other normal galaxies, which enable us to model
cosmic-ray intensities there, study the intensity evolution and its
dependence on the supernova rate, gas density etc., and, therefore, to
understand the \emph{history} of cosmic rays in the Milky Way galaxy.
GLAST with its high sensitivity and
resolution should also provide a final proof of proton acceleration in
SNRs -- long awaited by the cosmic-ray community. This will provide
insight into the processes of acceleration of protons and electrons by
SNR shocks and shed light on the puzzle of the low $e/p$-ratio in cosmic rays.

Among other goals, new accurate measurements of cosmic-ray nuclei,
positrons, and antiprotons are desirable.  Produced in the same
$pp$-interactions as \grays\ and positrons,  antiprotons with their
unique spectral shape are seen as a key link between physics of cosmic
rays and diffuse \grays\ and could provide important clues to such
problems as Galactic cosmic-ray propagation, possible imprints of our
local environment, heliospheric modulation, dark matter etc. In a few
years, several high resolution space and balloon experiments are to be
launched. PAMELA (launch in 2004) is designed to measure antiprotons,
positrons, electrons, and isotopes H through C over the energy range
of 0.1 to 300 GeV. Future Antarctic flights of a new BESS-Polar
instrument will considerably increase the accuracy of data on
antiprotons and light elements. AMS will measure cosmic-ray particles
and nuclei
$Z\hbox{\rlap{\hbox{\lower3pt\hbox{$\sim$}}}\lower-2pt\hbox{$<$}}26$
from GeV to TeV energies. This is complemented by low energy missions,
ACE, Ulysses, and Voyager which will continue to deliver excellent
quality spectral and isotopic data $Z\leq28$, and TIGER capable of
measuring heavier nuclei $Z>29$.  Several missions are planned to
target specifically the high energy electron spectrum, which could
provide unique information about our local environment and sources of
cosmic rays nearby.

\acknowledgments

A part of this work has been done  during a visit of I.~Moskalenko to
the Max-Planck-Institut f\"ur extraterrestrische Physik in Garching;
the warm hospitality and financial support of the Gamma Ray Group is
gratefully acknowledged.  The work by I.~Moskalenko  was supported in
part by a NASA Astrophysics Theory Program grant.

\end{document}